\documentclass[journal]{IEEEtran}
\usepackage[cmex10]{amsmath}
\usepackage{graphicx}
\usepackage{url}
\usepackage{color}
\usepackage{algorithmicx}
\usepackage{algorithm}
\usepackage[noend]{algpseudocode}
\begin{document}

\title{Granger Causality in Multi-variate Time Series using a Time Ordered Restricted Vector Autoregressive Model}

\author{Elsa~Siggiridou and~Dimitris~Kugiumtzis%
\thanks{E. Siggiridou and D. Kugiumtzis are with the Department of Electrical and Computer Engineering, Aristotle University of Thessaloniki, Thessaloniki 54124, Greece, e-mail: esingiri@auth.gr, dkugiu@auth.gr}}

\maketitle

\begin{abstract}
Granger causality has been used for the investigation of the
inter-dependence structure of the underlying systems of
multi-variate time series. In particular, the direct causal
effects are commonly estimated by the conditional Granger
causality index (CGCI). In the presence of many observed variables
and relatively short time series, CGCI may fail because it is
based on vector autoregressive models (VAR) involving a large
number of coefficients to be estimated. In this work, the VAR is
restricted by a scheme that modifies the recently developed method
of backward-in-time selection (BTS) of the lagged variables and
the CGCI is combined with BTS. Further, the proposed approach is
compared favorably to other restricted VAR representations, such
as the top-down strategy, the bottom-up strategy, and the least
absolute shrinkage and selection operator (LASSO), in terms of
sensitivity and specificity of CGCI. This is shown by using
simulations of linear and nonlinear, low and high-dimensional
systems and different time series lengths. For nonlinear systems,
CGCI from the restricted VAR representations are compared with
analogous nonlinear causality indices. Further, CGCI in
conjunction with BTS and other restricted VAR representations is
applied to multi-channel scalp electroencephalogram (EEG)
recordings of epileptic patients containing epileptiform
discharges. CGCI on the restricted VAR, and BTS in particular,
could track the changes in brain connectivity before, during and
after epileptiform discharges, which was not possible using the
full VAR representation.
\end{abstract}

\begin{IEEEkeywords}
Granger causality, conditional Granger causality index (CGCI),
restricted or sparse VAR models, electroencephalogram
\end{IEEEkeywords}

% =========================================================
\IEEEpeerreviewmaketitle
\section{Introduction}
\label{intro}
% =========================================================

Granger causality has been applied to reveal inter-dependence
structure in multi-variate time series, first in econometrics
\cite{Luetkepohl05,Kirchgaassner07,Hoover08}, and then to other
areas and in particular to neuroscience, e.g. see the special
issue in \cite{Faes12}. According to the concept originally
introduced by Granger \cite{Granger69}, a variable $X$ Granger
causes another variable $Y$ if the prediction of $Y$ is improved
when $X$ is included in the prediction model for $Y$. In
multi-variate time series, the other observed variables are
included in the two vector autoregressive (VAR) models for $Y$.
The model including $X$ is called unrestricted or U-model, whereas
the one not including $X$ is called restricted or R-model. The
Granger causality is quantified by the conditional Granger
causality index (CGCI), defined as the logarithm of the ratio of
the error variances of the R-model and the U-model (the term
conditional stands for the general case of other observed
variables included in the two models) \cite{Geweke84,Guo08}. Apart
from CGCI and its formulations in the frequency domain, i.e. the
partial directed coherence \cite{Baccala01} and the direct
directed transfer function \cite{Korzeniewska03}, a number of
nonlinear Granger causality indices have been proposed based on
information theory \cite{Schreiber00,Kugiumtzis13a}, state space
dynamics \cite{Arnhold99,Chicharro09} and phase synchronization
\cite{Smirnov03} (for a comparison see \cite{Kreuz07,Papana13b}).
However, when inherently nonlinear and complex systems are studied
at small time intervals the nonlinear methods are not successfully
applicable \cite{Papana13b}. The same may hold for the CGCI and
other indices based on VAR models. In particular, the estimation
of the VAR coefficients may be problematic in the setting of many
observed variables and short time series.

The problem of reducing the number of model coefficients has been
addressed in linear multiple regression, and many methods of
variable subset selection have been developed. The optimal
solution is obtained by the computationally intensive and often
impractical search for all subset models. Suboptimal subset
selection methods include the simple sequential search method and
stepwise methods implementing the bottom-up (forward selection)
and top-down (backward elimination) strategies. More complicated
schemes have also been proposed, such as the genetic algorithms,
the particle swarm optimization and the ant colony optimizations.
The most popular methods seem to be the ridge regression and the
least absolute shrinkage and selection operator (LASSO), as well
as the combination of them, and other variants of LASSO
\cite{Miller02,Hastie09}. Other methods deal with dimension
reduction through transformation and projection of the variable
space, such as the principal component regression and the partial
least squares regression, or assume a smaller set of latent
variables, such as the dynamic factor models \cite{Stock06}. These
latter methods are not considered here because the purpose is to
identify the inter-dependence between the observed variables and
not between transformed or latent variables. All the
aforementioned methods are developed for regression problems and
do not take into account the lag dependence structure that
typically exists in multi-variate time series. This is addressed
in a recently developed method known as backward-in-time-selection
(BTS), which implements a supervised stepwise forward selection
guided by the lag order of the lagged variables \cite{Vlachos13}.
Starting the sequential selection from smallest to larger lags is
reasonable as in time series problems it is expected that the
variables at smaller lags are more explanatory to the response
variable (at the present time) than variables at larger lags. Thus
conditioning on the variables at smaller lags the variables at
larger lags may only enter the model if they have genuine
contribution not already contained in the selected variables of
smaller lags. It is reasonable to expect that in time series
problems the response (present) variable does not depend on all
lagged variables, as implied in the standard VAR approach. So,
dimension reduction may be useful in any case of multi-variate
time series, not only for the case of high dimensional time series
of short length, where VAR estimation fails.

So far there has been little focus on the problem of estimating
the linear Granger causality index in short time series of many
variables, where standard VAR model estimation may be problematic.
We are aware only of works attempting to estimate Granger
causality after restricting the VAR models using LASSO and
variants of this \cite{Arnold07,Lozano09,Shojaie10,He13,Basu15}
(for restricted VAR model using LASSO see also
\cite{Hsu08,Ren10,Ren13,Gefang14}). There are also applications of
LASSO in neuroscience, such as electroencephalograms
\cite{Haufe10} and fMRI \cite{Tang12,Pongrattanakul13}, and
bioinformatics, in microarrays \cite{Fujita07,Shojaie12}.

In this work, we propose a new dimension reduction approach
designed for time series. We modify and adapt the BTS method for
Granger causality, and develop the scheme called BTS-CGCI that
derives CGCI from the restricted VAR representation formed by BTS
(BTS-CGCI). Further, we compare BTS-CGCI to the CGCI derived by
other restricted VAR representations (top-down, bottom-up, LASSO)
and the full VAR representation. The ability of the different VAR
representations to identify the connectivity structure of the
underlying system to the observed time series is assessed by means
of Monte Carlo simulations. In the simulation study, we included
known linear and nonlinear, low and high-dimensional systems, and
considered different time series lengths. In particular, for
nonlinear systems, the comparison includes also two nonlinear
information based measures, i.e. the partial transfer entropy
\cite{Vakorin09,Papana12}, in analogy to CGCI from the full VAR
representation, and the partial mutual information from mixed
embedding (PMIME), in analogy to CGCI from the restricted VAR
\cite{Kugiumtzis13a}. Moreover, BTS-CGCI along with other
restricted and full VAR representations are applied to
multi-channel scalp electroencephalogram (EEG) recordings of
epileptic patients containing epileptiform discharges. The
obtained brain network before, during and after epileptiform
discharges, is assessed for each method.

The structure of the paper is as follows. In
Section~\ref{sec:BTSCGCI}, BTS-CGCI is presented along with other
VAR restriction methods as well as their statistical evaluation.
In Section~\ref{sec:Simulations}, the simulation study is
described and the results are presented. The application to EEG is
presented and discussed in Section~\ref{sec:RealData}. Finally,
the method and results are discussed in
Section~\ref{sec:Discussion}.

%%%%%%%%%%%%%%%%%%%%%%%%%%%%%%%%%%%%%%%%%%%%%%%%%%%%%%%%%%%%%%%%%%%%%%%
\section{Methods}
\label{sec:BTSCGCI}
%%%%%%%%%%%%%%%%%%%%%%%%%%%%%%%%%%%%%%%%%%%%%%%%%%%%%%%%%%%%%%%%%%%%%%%

% --------------------------------------------------
\subsection{The conditional Granger causality index}
% --------------------------------------------------

Let $X_t=\{X_{1,t},X_{2,t},\ldots,X_{K,t}\}$, $t=1,\ldots,N$, be a
$K$-dimensional stationary time series of length $N$. The
definition of the conditional Granger causality index (CGCI) from
a driving variable $X_i$ to a response variable $X_j$ involves two
vector autoregressive (VAR) models for $X_j$, called also dynamic
regression models\footnote{The VAR model implies a model for each
of the $K$ variables, but it is often used in the literature also
when the model regards only one variable (here $X_j$), so
alternatively and interchangeably we use the term dynamic
regression for the model of one variable.} \cite{Pankratz91}. The
first model is the unrestricted model (U-model) \cite{Brandt07},
given as
\begin{equation}
 X_{j,t}=\sum_{k=1}^K (a_{jk,1} X_{k,t-1}+\ldots+a_{jk,p} X_{k,t-p})+u_{j,t}
\label{eq:Umodel}
\end{equation}
where $p$ is the model order and $a_{jk,l}$ ($k=1,\ldots,K$,
$l=1,\ldots,p$) are the U-model coefficients. The U-model includes
all the $K$ lagged variables for lags up to the order $p$. The
second model is the restricted one (R-model) derived from the
U-model by excluding the lags of $X_i$, given as
\begin{equation}
X_{j,t}=\sum_{k=1,k\neq i}^K (b_{jk,1} X_{k,t-1}+\ldots+b_{jk,p}
X_{k,t-p})+e_{j,t} \label{eq:Rmodel}
\end{equation}
where $b_{jk,l}$ ($k=1,\ldots,K$ but $k\neq i$ and $l=1,\ldots,p$)
are the coefficients of the R-model. The terms $u_{j,t}$  and
$e_{j,t}$ are white noise with mean zero and variances
$\sigma^{2}_{U}$ and $\sigma^{2}_{R}$, respectively\footnote{In
general, $u_{j,t}$ and $e_{j,t}$ can be instantaneously correlated
with respect to $j=1,\ldots,K$, which then implies instantaneous
causal effects among the variables \cite{Luetkepohl05}, but this
setting is out of the scope of this work.}. Moreover, for
inference (see the significance test below) they are assumed to
have normal distribution. Fitting the U-model and R-model,
typically with ordinary least squares (OLS), we get the estimates
of the residual variances $\hat{\sigma}^{2}_{U}$ and
$\hat{\sigma}^{2}_{R}$. Then CGCI from $X_i$ to $X_j$ is defined
as
\begin{equation}
\mbox{CGCI}_{X_i \rightarrow X_j} = \ln
\frac{\hat{\sigma}^{2}_{R}}{\hat{\sigma}^{2}_{U}}. \label{eq:CGCI}
\end{equation}
CGCI is at the zero level when $X_i$ does not improve the
prediction of $X_j$ (the U-model and R-model give about the same
fitting error variance) and obtains larger positive values when
$X_i$ improves the prediction of $X_j$ indicating that $X_i$
Granger causes $X_j$.

The statistical significance of CGCI is commonly assessed by a
parametric significance test on the coefficients of the lagged
driving variable $X_i$ in the U-model \cite{Brandt07}. The null
hypothesis is H$_0$: $a_{ji,l}=0$ for all $l=1,\ldots,p$, and the
Fisher statistic is
\begin{equation}
F =\frac{(\mbox{SSE}^R-\mbox{SSE}^U)/p}{\mbox{SSE}^U/((N-p)-Kp)},
\label{eq:Fstat}
\end{equation}
where SSE is the sum of squared errors and the superscript denotes
the model, $N-p$ is the number of equations and $Kp$ is the number
of coefficients of the U-model. The Fisher test assumes
independence of observations, normality and equal variance for the
observed variables, which may not be met in practice. We refrain
from discussing these issues here as the test is commonly applied
in the VAR based estimation of Granger causality, and only note
that the assumption of independence is likely to be violated as
the time series are typically auto- and cross- correlated. When
there are many observed variables (large $K$) and short time
series (small $N$), so that $Kp$ is large compared to $N$, the
estimation of the model coefficients may not be accurate and then
CGCI cannot be trusted.

% -----------------------------------------------------------------------------
\subsection{The modified backward in time selection}
\label{subsec:mBTS}
% -----------------------------------------------------------------------------

The backward-in-time-selection (BTS) method is a bottom-up
strategy designed for multi-variate time series to reduce the
terms in the VAR model in (\ref{eq:Umodel}) \cite{Vlachos13}. The
BTS method is developed so as to take into account feedback and
multi-collinearity, which are often observed in multi-variate time
series. The rationale is to select the order $p_k$ for each $X_k$
in the dynamic regression model in (\ref{eq:Umodel}) (instead of
having the same order $p$ for all variables) based on the inherent
property of time series that the dependence structure is closely
related to the temporal order of the variables. Thus BTS evaluates
progressively the inclusion of the lagged variables in the model,
starting with the most current lagged variables and moving
backwards in time. While the model derived by the BTS method in
\cite{Vlachos13} includes all lags up to the selected order $p_k$
for each $X_k$, here we modify the BTS method so as to include in
the model only the lags of each $X_k$ that are selected at each
step of the algorithm. The modified algorithm of BTS (mBTS) is
briefly presented below (we use the notation mBTS for the modified
algorithm but keep the notation BTS for the derived model and the
Granger causality approach). We note that the algorithm described
below is for determining one dynamic regression model for one of
the $K$ variables, say $X_j$, and it is repeated $K$ times for all
$K$ variables.

First, a maximum order $p_{\mbox{\footnotesize{max}}}$ is set
determining the maximum lag the algorithm will search for all $K$
variables to add in the dynamic regression model for the response
variable $X_j$. Thus the vector of all lagged variables that are
candidates to be included in the BTS model is
\[
\mathbf{W} = [X_{1,t-1} \ldots
X_{1,t-p_{\mbox{\footnotesize{max}}}} X_{2,t-1} \ldots
X_{K,t-p_{\mbox{\footnotesize{max}}}}],
\]
and has $Kp_{\mbox{\footnotesize{max}}}$ components. The algorithm
aims at finding an explanatory vector $\mathbf{w}_j$ formed from
the the most significant lagged variables of $\mathbf{W}$ in
predicting $X_j$, which thus comprise the BTS model. The
explanatory vector $\mathbf{w}_j$ is built progressively adding
one lagged variable at each cycle. The models formed by the
candidate explanatory vectors at each cycle are assessed with the
Bayesian information criterion (BIC) \cite{Schwarz78}. The steps
of the mBTS algorithm are given below and the pseudo-code is given
in Algorithm~\ref{mBTSalgorithm}.
%%%%%%%%%%%%% Algorithm 1 %%%%%%%%%%%%%%%%%%%
\begin{algorithm}
\caption{mBTS for $X_j$}
 \label{mBTSalgorithm}
\begin{algorithmic}[1]
% \Procedure{MyProcedure}{}
 \Require $X$: The set of $K$ time series
 \State $w \gets \emptyset$ \Comment{initially explanatory vector is
 empty}
 \State $\textit{BICold} \gets s^2$ \Comment{$s^2$: the error
 variance}
 \State $\textit{maxlags} \gets [0, \ldots, 0]$ \Comment{the $K$ maximum lags are
 initially set to zero}
 \While {$\textit{sum(maxlags)} < K \textit{pmax} $}
   \For {$i=1:K$}
        \If {$\textit{maxlags}(i) < \textit{pmax}$}
            \State $\textit{wcand} \gets \{w; (i,\textit{maxlags}(i)+1) \}$
            \State $\textit{BIC}(i) \gets
            \mbox{modelfit}(j,X,\textit{wcand})$
        \Else
            \State $\textit{BIC}(i) \gets \textit{BICold}$
        \EndIf
   \EndFor
   \State $[\textit{BICnew},k] \gets \min(\textit{BIC})$ \Comment{$k$ is the corresponding index to min}
   \If {$\textit{BICnew} < \textit{BICold}$}
        \State $\textit{BICold} \gets \textit{BICnew}$
        \State $w \gets \{w; (k,\textit{maxlags}(k)+1)\}$
        \State $\textit{maxlags}(k) \gets \textit{maxlags}(k) + 1$
    \Else
        \For {$i=1:K$}
            \State $\textit{maxlags}(i) \gets
            \min(\textit{maxlags}(i)+1, \textit{pmax})$
        \EndFor
    \EndIf
 \EndWhile
\Return $w$
\end{algorithmic}
\end{algorithm}
%%%%%%%%%%%%%%%%%%%%%%%%%%%%%%%%%%%%%%%%%%%%%%%%%%%%%%%%

\begin{enumerate}
    \item Initially, let the explanatory vector of lagged variables
    $\mathbf{w}_j^0$ be empty, corresponding to the zero-order
    model and BIC here is equal to the variance of $X_j$
    (lines 1-3
    in the pseudo-code; 'maxlags' is an auxiliary array keeping track of
    the maximum lag of each variable already searched, and therefore it is
    initially set to zero for each of the $K$ variables).
    \item For each of the $K$ variables separately,
    its largest tested lag so far is increased by one and this
    lagged variable is added to the current explanatory vector
    $\mathbf{w}_j^l$ (the superscript
    denotes the size of the current explanatory vector). In this way,
    $K$ candidate explanatory vectors are formed. For the first cycle,
    the current explanatory vector  $\mathbf{w}_j^0$ is empty and the
    $K$ candidate explanatory vectors are actually the $K$ scalar variables
\[
X_{1,t-1},\ldots, X_{K,t-1}.
\]
For the $l+1$ cycle, the explanatory vector $\mathbf{w}_j^l$ has
$l$ lagged variables, and let the maximum lag for each variable
$X_k$ in $\mathbf{w}_j^l$ be $\tau(l_k)$ ($\tau(l_k)$ is zero if
$X_k$ is not yet present in $\mathbf{w}_j^l$). Then the $K$
candidate explanatory vectors are \[ [\mathbf{w}_j^l \,\,
X_{1,t-\tau(l_1)-1}], \ldots, [\mathbf{w}_j^l \,\,
X_{K,t-\tau(l_K)-1}].
\]
Compute BIC for the $K$ dynamic regression models formed by the
$K$ candidate explanatory vectors (lines 5-10 in the pseudo-code;
the components of the candidate explanatory vector 'wcand' are
pairs of two numbers, the first indicating the variable and the
second the lag; the function 'modelfit' in line 8 provides the BIC
value for the fitted model defined by 'wcand').
    \item Find the candidate explanatory vector for which the
BIC value is smaller than the BIC value for $\mathbf{w}_j^l$. If
there are more than one such vectors (models) select the one with
the smallest BIC value. Update the current explanatory vector with
the explanatory vector of smallest BIC, denoted
$\mathbf{w}_j^{l+1}$ (lines 11-15 in the pseudo-code). The $l+1$
cycle is thus completed and go to the next step. If none of the
$K$ candidate explanatory vectors gives BIC smaller than the one
for $\mathbf{w}_j^l$, then increase the maximum lag $\tau(l_k)$
for each $X_k$ by one and go to the next step (if
$\tau(l_k)=p_{\mbox{\footnotesize{max}}}$ for any $k=1,\ldots,K$,
leave it as is; lines 16-18 in the pseudo-code).
    \item If all the maximum lags reached the maximum order,
    $\tau(l_k)=p_{\mbox{\footnotesize{max}}}$ ($k=1,\ldots,K$)
    then terminate, otherwise go to step 2 (line 4
    in the pseudo-code).
\end{enumerate}

Upon termination after $P_j$ cycles, the algorithm gives the final
explanatory vector $\mathbf{w}_j$ of size $P_j$ for $X_j$. It is
noted that $\mathbf{w}_j$ may not have lagged components of all
$K$ variables. Moreover, if a variable $X_k$ is present in
$\mathbf{w}_j$ with maximum lag $\tau(p_k)$, $\mathbf{w}_j$ may
not contain all lags of $X_k$ from one to $\tau(p_k)$, but a
subset of $p_k$ lags, and in general $p_k \le \tau(p_k)$ holds.
The latter constitutes a modification to the original BTS
algorithm in \cite{Vlachos13}, where for each $X_k$ in
$\mathbf{w}_j$ all lags up to the maximum lag $\tau(p_k)$ are
included in $\mathbf{w}_j$ and subsequently in the model for
$X_j$. The reason for this modification is to include in the
dynamic regression model for $X_j$ only terms (lagged variables)
found to improve the prediction of $X_j$, and this allows for the
identification of the exact lags of the variable that give
evidence of Granger causality. On the other hand, the inclusion of
all lags of a variable up to the selected order may possibly block
the presence of specific lags of other variables, and this would
reduce the sensitivity in detecting Granger causality effects.

\begin{figure*}[htb]
\centerline{\hbox{\includegraphics[width=6cm]{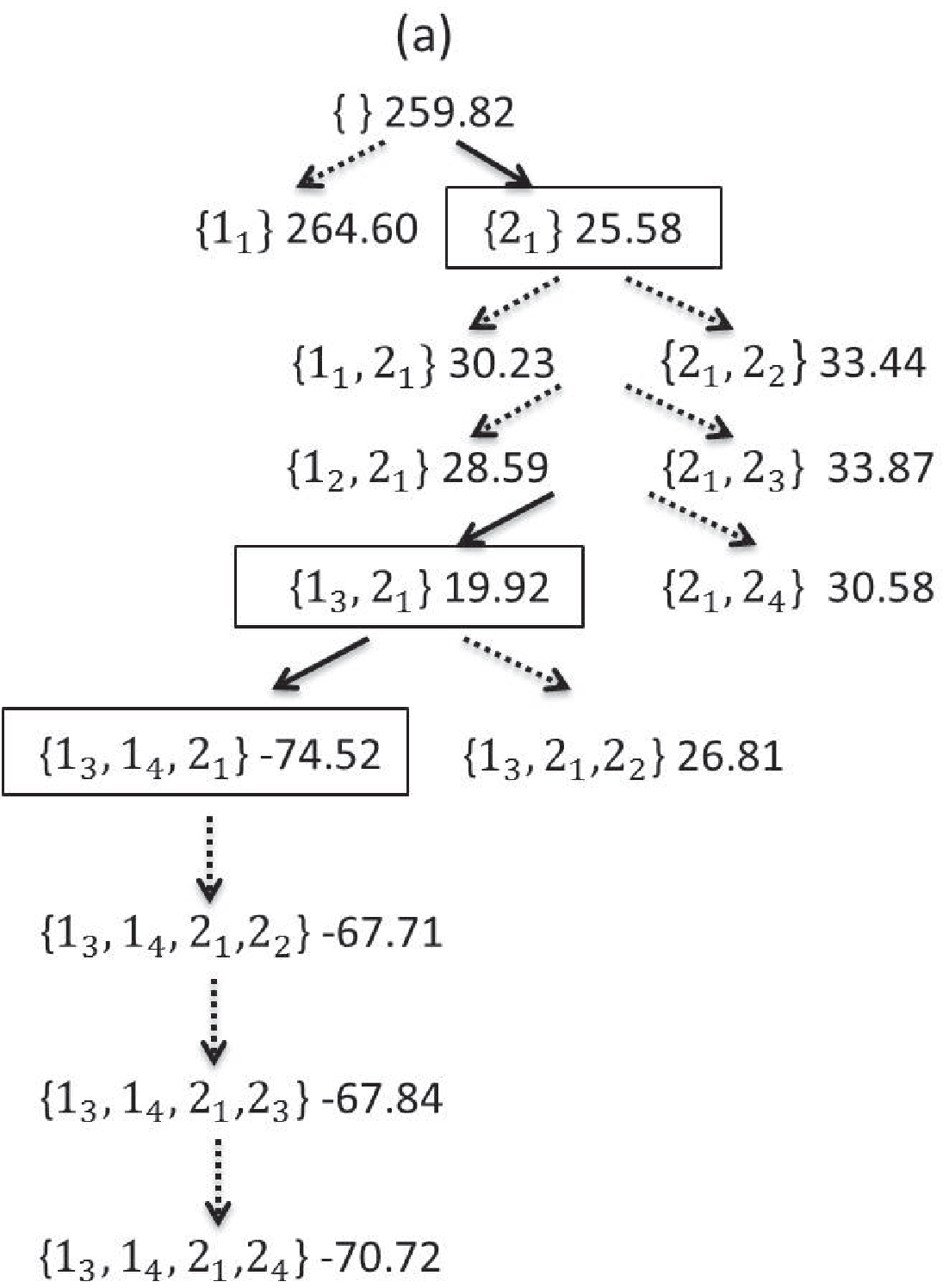}
\includegraphics[width=6cm]{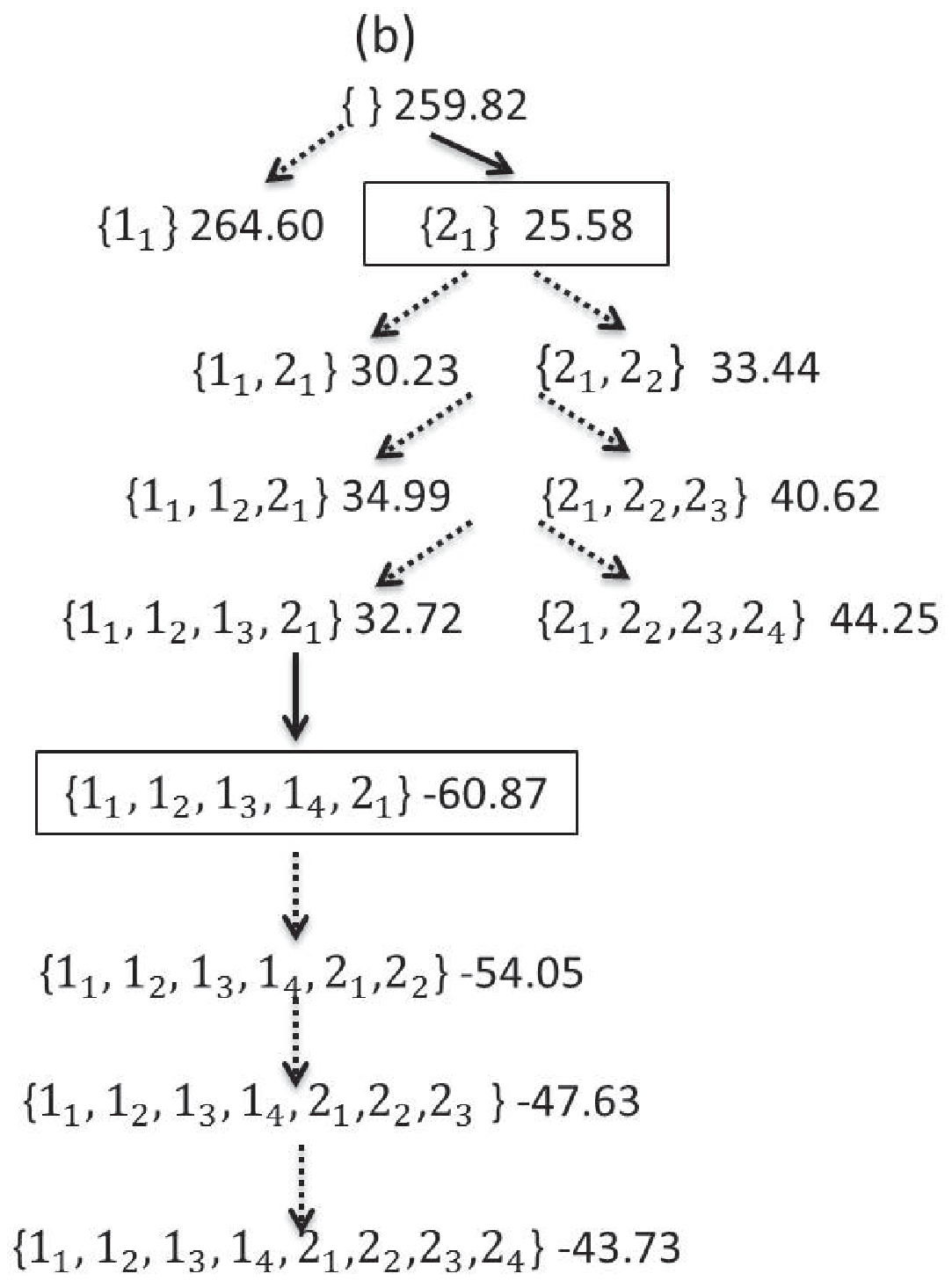}}}
\caption{(a) The steps of the mBTS algorithm for the second
response variable of the VAR(4) system in two variables. At each
step the candidate explanatory vector is shown (the integer number
indicates the variable and the subscript the lag) together with
the corresponding BIC value. When a lagged variable is selected
the vector and BIC value are shown in box and the arrow is given
with solid line (dashed line otherwise). (b) The same for BTS.}
\label{fig:stepexample}
\end{figure*}

The steps of mBTS are illustrated in Figure~\ref{fig:stepexample}
for a simple example of the following VAR(4) system in two
variables
\begin{eqnarray}
  \label{eq:BivariateSystem}
  X_{1,t} & = & 0.4X_{1,t-1}  + u_{1,t} \nonumber \\
  X_{2,t} & = & 0.4X_{2,t-1} - 0.3X_{1,t-4} + u_{2,t}
\end{eqnarray}
The explanatory vector $\mathbf{w}_2$ found by mBTS in
Figure~\ref{fig:stepexample}a is different than that found by BTS
in Figure~\ref{fig:stepexample}b. The explanatory vector by mBTS
includes the two true lagged variables and one other lagged
variable $X_{1,t-3}$ not included in the expression for $X_{2,t}$,
which however does not result in wrong causal effects, while BTS
includes also two more lagged variables, i.e. all lagged variables
of $X_1$ up to order four.
%----------------
%FIGURE(1)
%----------------

% -----------------------------------------------------------------------------
\subsection{The conditional Granger causality index and backward-in-time selection}
\label{subsec:CGCIBTS}
% -----------------------------------------------------------------------------

We propose to calculate $\mbox{CGCI}_{X_i \rightarrow X_j}$ making
use of the mBTS algorithm on the lagged variables of
\{$X_{1}$,$X_{2}$,...,$X_{K}$\} in the following way. The U-model
for $X_j$ is the dynamic regression model formed by
$\mathbf{w}_j$. If none of the lagged variables of $X_i$ is
present in $\mathbf{w}_j$ and subsequently in the U-model, then
$\mbox{CGCI}_{X_i \rightarrow X_j}$ is zero (the U-model and
R-model are identical). Otherwise the structure of the R-model is
formed from that of the U-model dropping all the lagged $X_i$
components. Specifically, we consider the following representation
of the explanatory vector $\mathbf{w}_j$
\[
\mathbf{w}_j = [\mathbf{w}_{j,1} \,\, \mathbf{w}_{j,2} \,\, \ldots
\mathbf{w}_{j,K}]
\]
meaning that $\mathbf{w}_j$ is decomposed to vectors of lags from
each variable $X_k$. For each $k=1,\ldots,K$, $\mathbf{w}_{j,k}$
is
\[
\mathbf{w}_{j,k} = \{X_{k,t-\tau(1)},\ldots,X_{k,t-\tau(p_k)} \},
\]
and $\tau(l)$ ($l=1,\ldots,p_k$), denote the $p_k$ selected lags
of variable $X_k$ by mBTS. The length of $\mathbf{w}_j$ is
$P_j=\sum_{k=1}^K p_k$. The U-model from mBTS for $X_j$ is
\begin{equation}
 X_{j,t}=\sum_{k=1}^K \mathbf{a}_{j,k} \mathbf{w}_{j,k}^T +u_{j,t}
\label{eq:BTSUmodel}
\end{equation}
where $\mathbf{a}_{j,k}$ is a row vector of $p_k$ coefficients and
the transpose $^T$ sets $\mathbf{w}_{j,k}$ in column form. The
R-model with respect to the causality effect $X_i \rightarrow X_j$
is
\begin{equation}
 X_{j,t}=\sum_{k=1,k\neq i}^K \mathbf{b}_{j,k} \mathbf{w}_{j,k}^T +e_{j,t}
\label{eq:BTSRmodel}
\end{equation}
where $\mathbf{b}_{j,k}$ is a row vector of coefficients as
$\mathbf{a}_{j,k}$. The two models are fitted to the time series
with OLS and the CGCI is computed as for the full VAR in
(\ref{eq:CGCI}) from the residuals of the U-model and the R-model.
Obviously, if mBTS does not find any explanatory lags of $X_i$,
i.e. $\mathbf{w}_{j,i}$ is empty, then the U-model and R-model are
identical and $\mbox{CGCI}_{X_i \rightarrow X_j}=0$, whereas if
$\mathbf{w}_{j,i}$ has at least one lagged component of $X_i$ then
$\mbox{CGCI}_{X_i \rightarrow X_j}$ is positive. The last result
can be further tested for significance using the same parametric
test as for the full VAR but adapting the degrees of freedom in
the expression of the Fisher statistic as follows
\begin{equation}
F=\frac{(\mbox{SSE}^R-\mbox{SSE}^U)/p_i}
{\mbox{SSE}^U/((N-c)-P_j)}. \label{eq:Fstatrestricted}
\end{equation}
The parameters in (\ref{eq:Fstatrestricted}) are defined as
follows: $p_i$ is the number of lagged components of $X_i$ in the
U-model from mBTS for $X_j$ (replacing the VAR order $p$ in
(\ref{eq:Fstat})), $c=\max \{\tau_{p_k}\}$ is the largest lag in
the U-model (replacing $p$ in (\ref{eq:Fstat})), $N-c$ determines
the number of equations for OLS estimation of the model
coefficients, and $P_j$ is the total number of the U-model
coefficients (replacing $Kp$ in (\ref{eq:Fstat})).

We note that the R-model is formed by omitting the lags of the
driving variable $X_i$ from the U-model for the response variable
$X_j$ obtained by mBTS. This is not exactly equivalent to the
standard Granger causality approach using the full VAR in that the
structure of the R-model without the $X_i$ components is obtained
directly from the U-model (under mBTS) and it is not optimized.
The optimization here would require running mBTS again for all but
the $X_i$ lagged variables. This approach has the computational
cost of applying mBTS $K-1$ times (for each driving variable
$i=1,\ldots,K$ and $i\neq j$). Moreover, the significance test
using the Fisher statistic in (\ref{eq:Fstatrestricted}) could not
be applied as the terms of the R-model would not in general
constitute a subset of the set of the terms of the U-model. We
thus opted in adapting the R-model in (\ref{eq:BTSRmodel}).

It is noted that having a U-model of different terms (determined
by mBTS) for each response variable $X_j$ makes the CGCI values
across the response variables not directly comparable. This can be
seen from the different degrees of freedom of the Fisher
distribution of the test statistic (different $p_i$ and $P_j$ in
(\ref{eq:Fstatrestricted})), which suggests different critical
values for the statistical significance of the causal effects.
However, in practice the differences in the significance level of
CGCI are not substantial, as demonstrated in the simulation study,
e.g. see Figure~\ref{fig:boxplots}. Note that this holds for any
dimension reduction method.

% -----------------------------------------------------------------------------
\subsection{Other VAR restriction schemes}
\label{sec:OtherMethods}
% -----------------------------------------------------------------------------

The other methods restricting the VAR model considered for
comparison to BTS in this study are the top-down strategy, the
bottom-up strategy and LASSO. For $K$ variables, maximum lag
$p_{\mbox{\footnotesize{max}}}$ and a response variable $X_j$, all
methods attempt to constrain the dynamic regression model in
(\ref{eq:Umodel}).

The top-down strategy starts with the full model in
(\ref{eq:Umodel}), and at each step checks whether dropping a term
attains a lower BIC value. If it does, the term is removed from
the model otherwise it is retained in the model, and the same
procedure is repeated for the next term. The order of the tested
terms can be arbitrary, and in \cite{Luetkepohl05} the order is
first for the largest lag $p_{\mbox{\footnotesize{max}}}$
searching for all variables starting from $X_K$ down to $X_1$,
then for the next largest lag $p_{\mbox{\footnotesize{max}}-1}$
and so on. We call this scheme TDlag. For the example of the
bivariate system in (\ref{eq:BivariateSystem}), the TDlag starts
with all 8 lagged variables ($K=2$,
$p_{\mbox{\footnotesize{max}}}=4$), removes first $X_{2,t-4}$,
retains $X_{1,t-4}$, removes all lagged variables down to
$X_{2,t-1}$, which is retained, and finally removes $X_{1,t-1}$,
giving the true explanatory vector $\mathbf{w}_2 = [X_{1,t-4},
X_{2,t-1}]$. We also consider the scheme inter-changing the order
of lags and variables, searching first for all lags of $X_K$ from
$p_{\mbox{\footnotesize{max}}}$ down to 1, then the same for
$X_{K-1}$ and so on, and we call this scheme TDvar.

On the contrary, the bottom-up strategy builds up the model adding
instead of removing terms \cite{Luetkepohl05}. The order for
adding terms is again arbitrary. It starts with the best model
regressing $X_j$ only on $X_1$ of an order $p_1$ determined by BIC
(all lagged terms of $X_1$ up to $p_1$ are included). Then the
same is applied for $X_2$ conditioned on the $p_1$ terms of $X_1$,
so that $p_2$ terms of $X_2$ are added to the model ($p_2$ can
also be zero), and so on. To account for over-estimation of the
lagged variables, the top-down procedure is then applied to the
derived model. This scheme is called BUlag if we apply TDlag to
the derived model. For the example in (\ref{eq:BivariateSystem}),
the BUlag finds first the best model of $X_2$ only on $X_1$ to be
of order four and then this model is augmented only with the first
lag of $X_2$, including thus five terms. Then applying TDlag the
three terms are removed and the final explanatory vector
$\mathbf{w}_2 = [X_{1,t-4}, X_{2,t-1}]$ is the true one. We also
consider the search first along the variables and then lags of the
derived model (TDvar), and we call this scheme BUvar.

The method of least absolute shrinkage and selection operator
(LASSO) is a least squares method with $L_1$ constraint on the
regression parameters. LASSO is used here for the estimation of
the coefficients of the lagged variables in the dynamic regression
model (1) for $X_j$ stacked in vector form,
$\mathbf{a}=\{a_{jk,l}\}$, obtained as
\begin{align*}
\hat{\mathbf{a}} = & \arg\min_{\mathbf{a}} \{
\sum\limits_{t=p_{\mbox{\scriptsize{max}}}+1}^N
(X_{j,t}-\sum_{l=1}^{p_{\mbox{\scriptsize{max}}}} \sum_{k=1}^K
a_{jk,l} X_{k,t-l})^2\} \\
& \mbox{subject to} \,
\sum\limits_{l=1}^{p_{\mbox{\scriptsize{max}}}}
\sum\limits_{k=1}^K \vert a_{jk,l}\vert <s,
\end{align*}
where $s$ is a tuning parameter \cite{Songsiri10,Avventi13}. The
so-called covariance test \cite{Lockhart14} is applied in order to
find the statistically significant $a_{jk,l}$. As implemented
here, first $s$ is increased and the terms (lagged variables)
entering the model are tracked (we use the 'lars' function in R
language \cite{Rlanguage08}). Then the statistically significant
terms are found testing them at the reverse order of appearance in
the model (we use the 'covTest' function in R language
\cite{Rlanguage08}). For the example in
(\ref{eq:BivariateSystem}), the lagged terms entering the model
are found by 'lars' at 8 steps: first $X_{2,t-1}$, then
$X_{1,t-4}$, and then the following 6 terms. Then the covariance
test runs sequentially and it removes the six last terms not found
statistically significant (at the $\alpha=0.05$ level), and
retains $X_{1,t-4}$ and then $X_{2,t-1}$ found statistically
significant, giving finally the true explanatory vector
$\mathbf{w}_2 = [X_{1,t-4}, X_{2,t-1}]$.

All these approaches give the structure of the U-model and we
consider the R-model by omitting the lags of the driving variable
and further compute CGCI, as described for the BTS approach in
Sec.~\ref{subsec:mBTS}.

% -----------------------------------------------------------------------------
\subsection{Statistical evaluation of method accuracy}
% -----------------------------------------------------------------------------

For a system of $K$ variables there are $K(K-1)$ ordered pairs of
variables to search for causality. The significance test of CGCI
either in the case of the full VAR model (see (\ref{eq:Fstat})) or
the restricted VAR model (see (\ref{eq:Fstatrestricted})) is thus
applied for each of the $K(K-1)$ pairs. To correct for multiple
testing, the false discovery rate (FDR) can be used
\cite{Benjamini95}, briefly presented below. The $p$-values of the
$m=K(K-1)$ significance tests are set in ascending order $p_{(1)}
\le p_{(2)} \le \ldots \le p_{(m)}$. The rejection of the null
hypothesis of zero CGCI at the significance level $\alpha$ is
decided for all variable pairs for which the $p$-value of the
corresponding test is less than $p_{(k)}$, where $p_{(k)}$ is the
largest $p$-value for which $p_{(k)} \le k \alpha /m$ holds. We
adopt the FDR correction to determine the presence of Granger
causality for all the ordered pairs of variables of the systems we
study below.

In the simulations of known systems, we know the true coupling
pairs and thus we can compute performance indices for rating the
methods of causality. Here we consider the specificity,
sensitivity, Matthews correlation coefficient, F-measure and
Hamming distance.

The sensitivity is the proportion of the true causal effects (true
positives, TP) correctly identified as such, given as TP/(TP+FN),
where FN (false negatives) denotes the number of pairs having true
causal effects but have gone undetected. The specificity is the
proportion of the pairs correctly not identified as having causal
effects (true negatives, TN), given as TN/(TN+FP), where FP (false
positives) denotes the number of pairs found falsely to have
causal effects. An ideal method would give values of sensitivity
and specificity at one. However, this is seldom attainable and in
order to weigh sensitivity and specificity collectively we
consider the Matthews correlation coefficient (MCC)
\cite{Matthews75}, given as
\[
 \text{MCC}=\frac{\text{TP}*\text{TN}-\text{FP}*\text{FN}}
{\sqrt{\text{(\text{TP}+\text{FP})*(\text{TP}+\text{FN})*(\text{TN}+\text{FP})*(\text{TN}+\text{FN})}}},
\]
where $*$ denotes multiplication. MCC ranges from -1 to 1. If
MCC=1 there is perfect identification of the pairs of true and no
causality, if MCC=-1 there is total disagreement and pairs of no
causality are identified as pairs of causality and vice versa,
whereas MCC at the zero level indicates random assignment of pairs
to causal and non-causal effects.

Similarly, we consider the F-measure that combines precision and
sensitivity. The precision, called also positive predictive value,
is the number of detected true causal effects divided by the total
number of detected casual effects, given as TP/(TP+FP), and  the
F-measure (FM) is defined as
\[
\mbox{FM} = \frac{2*\text{precision} *
\text{sensitivity}}{\text{precision} + \text{sensitivity}} =
\frac{2\text{TP}}{2\text{TP}+\text{FN}+\text{FP}},
\]
which ranges from 0 to 1. If FM=1 there is perfect identification
of the pairs of true causality, whereas if FM=0 no true coupling
is detected.

The Hamming distance (HD) is the sum of false positives (FP) and
false negatives (FN). Thus HD gets non-negative integer values
bounded below by zero (perfect identification) and above by
$K(K-1)$ if all pairs are misclassified.

To assess the statistical significance of the performance indices
between the proposed method of mBTS and the other restriction
schemes we conduct paired t-test for means, as suggested in
\cite{Demsar06}, with correction of multiple comparisons
(Bonferroni correction \cite{Dunn59}). We have confirmed the same
results using randomization test, i.e. shuffling randomly the
index values in the two groups (methods) to generate each
randomized copy of the paired sample.

%%%%%%%%%%%%%%%%%%%%%%%%%%%%%%
\section{Simulations}
\label{sec:Simulations}
%%%%%%%%%%%%%%%%%%%%%%%%%%%%%%

% -------------------------------
\subsection{The simulation setup}
% -------------------------------

\begin{figure*}[htb]
\centerline{\hbox{\includegraphics[width=43mm]{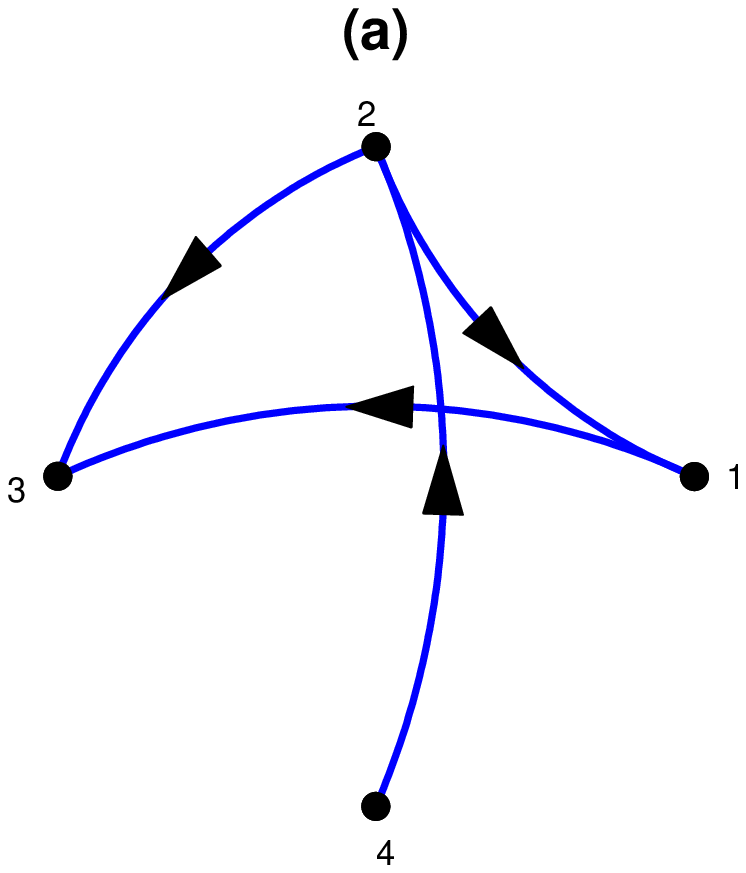}
\hspace{2mm} \includegraphics[width=43mm]{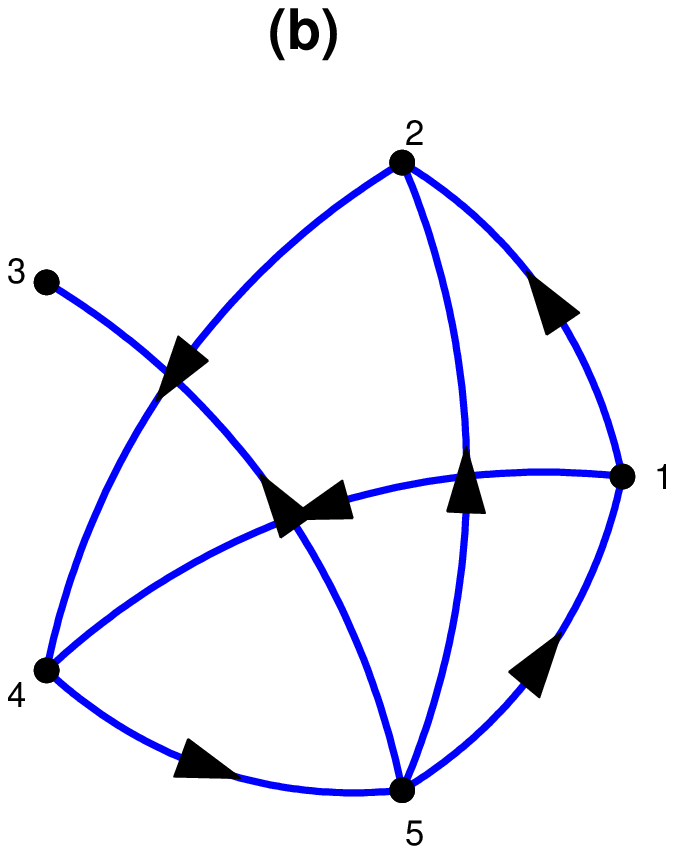}
\hspace{2mm} \includegraphics[width=43mm]{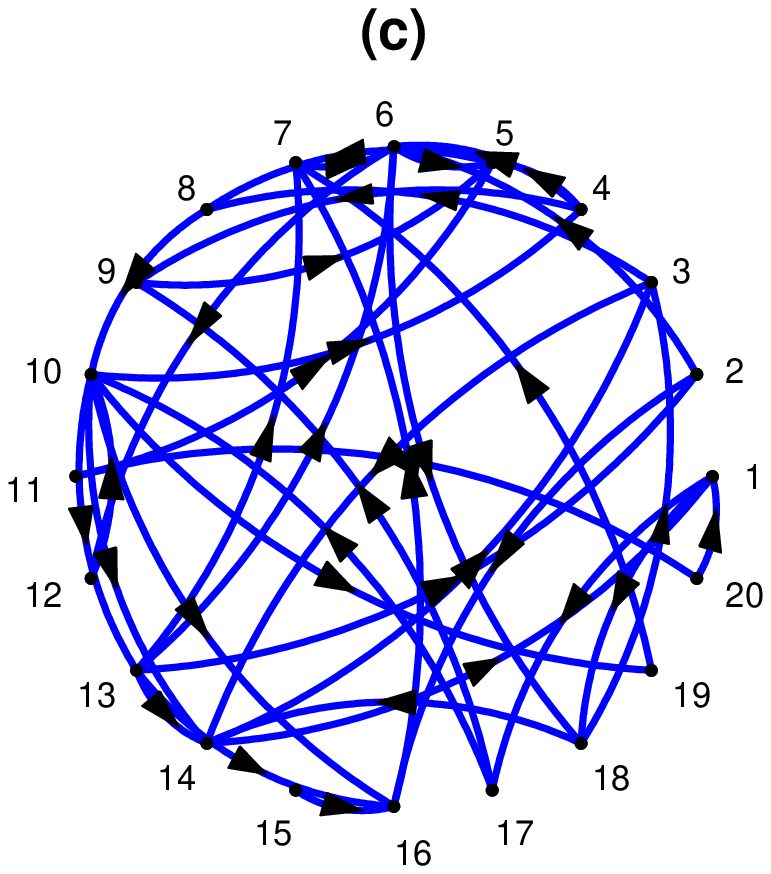}
\hspace{2mm} \includegraphics[width=43mm]{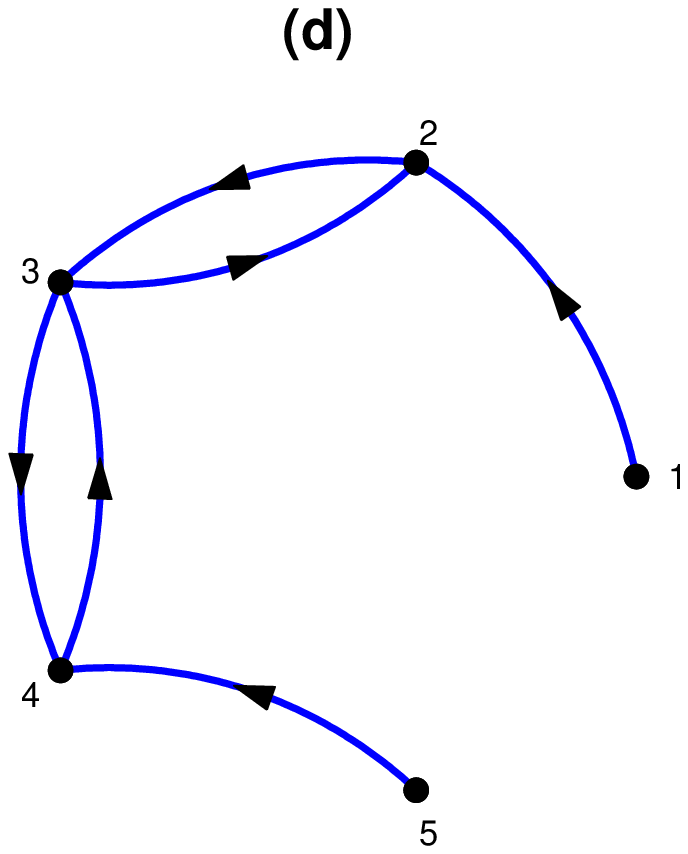}}}
\caption{The graphs of the connectivity structure of the simulated
systems: (a) S1, (b) S2, (c) S3, and (d) S4.}
\label{fig:connectgraphs}
\end{figure*}

In the simulations, we compare CGCI obtained by BTS to CGCI
obtained by the other VAR restriction methods, i.e. LASSO, the two
top-down strategies (TDlag and TDvar), the two bottom-up
strategies (BUlag and BUvar), as well as CGCI from the full VAR
model, denoted Full, on time series from known systems. The
simulation systems are as follows:

{\bf S1}: A VAR(4) process on $K=5$ variables (model 1 in Schelter
et al (2006) \cite{Schelter06e}),
\begin{eqnarray}
  \label{eq:Schelter}
  X_{1,t} & = & 0.4X_{1,t-1} - 0.5X_{1,t-2} + 0.4X_{5,t-1} + u_{1,t} \nonumber\\
  X_{2,t} & = & 0.4X_{2,t-1} - 0.3X_{1,t-4} + 0.4X_{5,t-2} + u_{2,t} \nonumber\\
  X_{3,t} & = & 0.5X_{3,t-1} - 0.7X_{3,t-2} - 0.3X_{5,t-3} + u_{3,t} \\
  X_{4,t} & = & 0.8X_{4,t-3} + 0.4X_{1,t-2} + 0.3X_{2,t-2} + u_{4,t} \nonumber\\
  X_{5,t} & = & 0.7X_{5,t-1} - 0.5X_{5,t-2} - 0.4X_{4,t-1} + u_{5,t} \nonumber
\end{eqnarray}
The connectivity structure of S1 is shown as graph in
Figure~\ref{fig:connectgraphs}a.

%----------------
%FIGURE(2)
%----------------

{\bf S2}: A VAR(5) process on $K=4$ variables (model 1 in
Winterhalder et al (2005)\cite{Winterhalder05}),
\begin{eqnarray}
  X_{1,t} & = & 0.8X_{1,t-1} + 0.65X_{2,t-4} + \epsilon_{1,t} \nonumber\\
  X_{2,t} & = & 0.6X_{2,t-1} + 0.6X_{4,t-5} + \epsilon_{2,t}  \nonumber\\
  X_{3,t} & = & 0.5X_{3,t-3} - 0.6X_{1,t-1} + 0.4X_{2,t-4} + \epsilon_{3,t} \nonumber\\
  X_{4,t} & = & 1.2X_{4,t-1} - 0.7X_{4,t-2} + \epsilon_{4,t} \nonumber
% \label{eq:Winterlander}
\end{eqnarray}
The connectivity structure of S2 is shown as graph in
Figure~\ref{fig:connectgraphs}b.

{\bf S3}: A VAR(3) process on $K=20$ as suggested in
\cite{Basu15}. Initially 10\% of the coefficients of VAR(3) are
set to one and the rest are zero and the positive coefficients are
reduced iteratively until the stationarity condition is fulfilled.
The autoregressive terms of lag one are set to one. The
connectivity structure of S3 is shown as graph in
Figure~\ref{fig:connectgraphs}c.

{\bf S4}: A nonlinear system, the H\'{e}non coupled maps of $K$
variables \cite{Politi92,Kugiumtzis13a}, where the first and last
variable in the chain of $K$ variables drive their adjacent
variable and the other variables drive the adjacent variable to
their left and right,
\begin{align*}
X_{i,t} = & 1.4-X_{i,t-1}^2+0.3X_{i,t-2}, \quad\quad \mbox{for} \,\, i=1,K  \\
X_{i,t} = & 1.4 - \left( 0.5 C (X_{i-1,t-1} + X_{i+1,t-1}) + (1-C)X_{i,t-1} \right)^2 \\
& + 0.3 X_{i,t-2}, \quad\quad  \mbox{for} \,\, i=2,\ldots,K-1
\end{align*}
Different number of variables $K$ are considered. The connectivity
structure of S4 for $K=5$ is shown as graph in
Figure~\ref{fig:connectgraphs}d.

The systems S1, S2 and S3 are used to test the sensitivity and
specificity of the methods, and the system S4 is used to assess
the usefulness of the proposed restricted CGCI also when the
multi-variate system is nonlinear. We make 1000 realizations for
each system for different orders $p_{\mbox{\footnotesize{max}}}$
and time series length $N$.

% ------------------------------------------------------
\subsection{An illustrative example for VAR restriction}
% ------------------------------------------------------

The performance of BTS as well as the VAR restriction methods
LASSO, TDlag and BUlag on the first variable of system S1 in
(\ref{eq:Schelter}) in identifying the true explanatory lagged
variables $X_{1,t-1}$, $X_{1,t-2}$ and $X_{5,t-1}$ is illustrated
in Table~\ref{tab:detailed} for $p_{\mbox{\footnotesize{max}}}=4$
and $N=100$.
%----------------
%TABLE(1)
%----------------
\begin{table*}
\centering
\renewcommand{\arraystretch}{1}
\caption{Average over 1000 realizations of the coefficients
estimated by the methods BTS, LASSO, TDlag and BUlag for the first
variable of S1 using $p_{\mbox{\footnotesize{max}}}=4$ and
$N=100$. Each cell presents the average parameter value and in
parentheses the frequency of occurrence of the lagged variable in
the model. The true lagged variables are highlighted in frame
boxes.} \label{tab:detailed}
\begin{tabular}{c||c c c c c}
\hline \hline
 & BTS  & LASSO  & TDlag  & BUlag   \\
 \hline \hline
\framebox[1\width]{$X_{1,t-1}$} &  0.275 (65.9\%)  & 0.159 (51.4\%)  & 0.433 (97.7\%)  & 0.463 (98.9\%)   \\
\framebox[1\width]{$X_{1,t-2}$} & -0.448 (99.9\%)  & -0.246 (58.2\%) & -0.501 (99.8\%)  & -0.514 (99.9\%)  \\

$X_{1,t-3}$ & -0.003 (5.1\%)  & -0.060 (21.6\%)   & -0.007 (5.8\%)   & 0.000 (1.4\%)    \\
$X_{1,t-4}$ & -0.002 (1.7\%) & -0.001 (0.1\%) & -0.003 (3.9\%) & -0.000 (1.4\%) \\

\hline
$X_{2,t-1}$ & -0.016 (22.3\%)  & 0.000 (0.1\%)   & -0.002 (5.9\%)   & -0.003 (3.5\%)    \\
$X_{2,t-2}$& -0.000 (1.5\%) & -0.001 (0.2\%)  & 0.002 (5.5\%) & -0.002 (0.9\%)  \\

$X_{2,t-3}$& 0.001 (0.9\%)   & 0.001 (0.5\%)   & 0.00 (7\%)   & -0.001 (0.4\%)   \\
$X_{2,t-4}$& -0.000 (0.4\%)     & -0.001 (0.7\%) & -0.009 (5.8\%) & -0.000 (0.1\%) \\

\hline
$X_{3,t-1}$ &  0.010 (16.4\%)  & 0.001 (1\%)   & 0.001 (7.2\%)   & 0.002 (3.4\%)   \\
$X_{3,t-2}$& -0.000 (3.2\%)  & -0.000 (0.4\%) & 0.002 (6.3\%)   & 0.001 (0.6\%)    \\

$X_{3,t-3}$& 0.001 (0.6\%) & -0.000 (0.8\%)   & 0.001 (4.8\%)   & -0.005 (0.3\%)    \\
$X_{3,t-4}$& -0.001 (0.2\%) & -0.000 (0.6\%) & -0.001 (6.5\%) & -0.000 (0.1\%)  \\
\hline
$X_{4,t-1}$& -0.009 (7.5\%)  & -0.000 (0.1\%)   & 0.006 (5.4\%)   & 0.004 (5.2\%)    \\
$X_{4,t-2}$& 0.001 (1\%)  & -0.000 (0.1\%)  & -0.004 (6.2\%) & -0.006 (2.4 \%)  \\

$X_{4,t-3}$& -0.000 (0.3\%)   & -0.000 (0.4\%)   & -0.005 (5.8\%)   & 0.000 (2.5\%)    \\
$X_{4,t-4}$& -0.000 (0.1\%)     & -0.000 (0.2\%) & -0.001 (5.6\%) & 0.001 (0.7\%) \\

\hline
\framebox[1\width]{$X_{5,t-1}$} & 0.424 (99.7\%)  & 0.379 (82.8\%)   & 0.174 (39\%)   &  0.114 (25.7\%)   \\
$X_{5,t-2}$& 0.012 (8.7\%)  & 0.011 (3.7\%) & 0.014 (7.2\%)  & 0.083 (3.6\%)  \\

$X_{5,t-3}$& -0.001 (2.9\%)  & -0.000 (0.1\%)   & -0.005 (6.3\%)   & -0.004 (0.6\%)    \\
$X_{5,t-4}$& -0.001 (0.7\%) & -0.007 (2.7\%) & -0.003 (6.3\%) & -0.000 (0.3\%)  \\

   \hline \hline
   \end{tabular}
   \end{table*}
All restriction methods reduce the full representation of 20
lagged variables to few lagged variables, most often being only
the true three lagged variables, so that the average values of the
coefficients from 1000 Monte Carlo realizations are close to zero
for all but the three true lagged variables. The false detection
is very low for TDlag and BUlag at about the 5\% significance
level (highest at 7.2\% and 5.2\%, respectively), and higher for
LASSO but at only one lagged variable (21.6\% for $X_{1,t-3}$),
and for BTS at two lagged variables (22.3\% and 16.4\% for
$X_{2,t-1}$ and $X_{3,t-1}$). On the other hand, BTS includes the
true lagged variables most often in the DR model, $X_{1,t-1}$ at
65.9\% of the runs and $X_{1,t-2}$ and $X_{5,t-1}$ almost always,
whereas the frequencies for LASSO are respectively 51.4\%, 58.2\%
and 82.8\%, and for TDlag and BUlag the first two are almost
always included while $X_{5,t-1}$ is included at a low rate of
39\% and 25.7\%, respectively. These results suggest that CGCI on
the basis of BTS would always detect the true driving effect $X_5
\rightarrow X_1$ (highest sensitivity) but also the false driving
effects $X_2 \rightarrow X_1$ and $X_3 \rightarrow X_1$ at a small
but significant percentage of cases (lower specificity), whereas
the other three methods would have high specificity (infrequently
detecting the driving effects from $X_2$, $X_3$ and $X_4$ to
$X_1$) but lower sensitivity as the true lagged variable
$X_{5,t-1}$ would not always enter the model.

% -------------------------------
\subsection{System 1}
% -------------------------------

The example above is for the dynamic regression of variable $X_1$
of S1, and in the following we present the Granger causality
effects for all variables of S1. The true Granger causality
effects between any variables of S1 are defined by the lagged
variables being present in the dynamic regression of each of the
five variables of S1 in (\ref{eq:Schelter}), being $X_1
\rightarrow X_2$, $X_1 \rightarrow X_4$, $X_2 \rightarrow X_4$,
$X_4 \rightarrow X_5$, $X_5 \rightarrow X_1$, $X_5 \rightarrow
X_2$ and $X_5 \rightarrow X_3$.

\begin{figure*}[htb]
\centerline{\includegraphics[width=16cm]{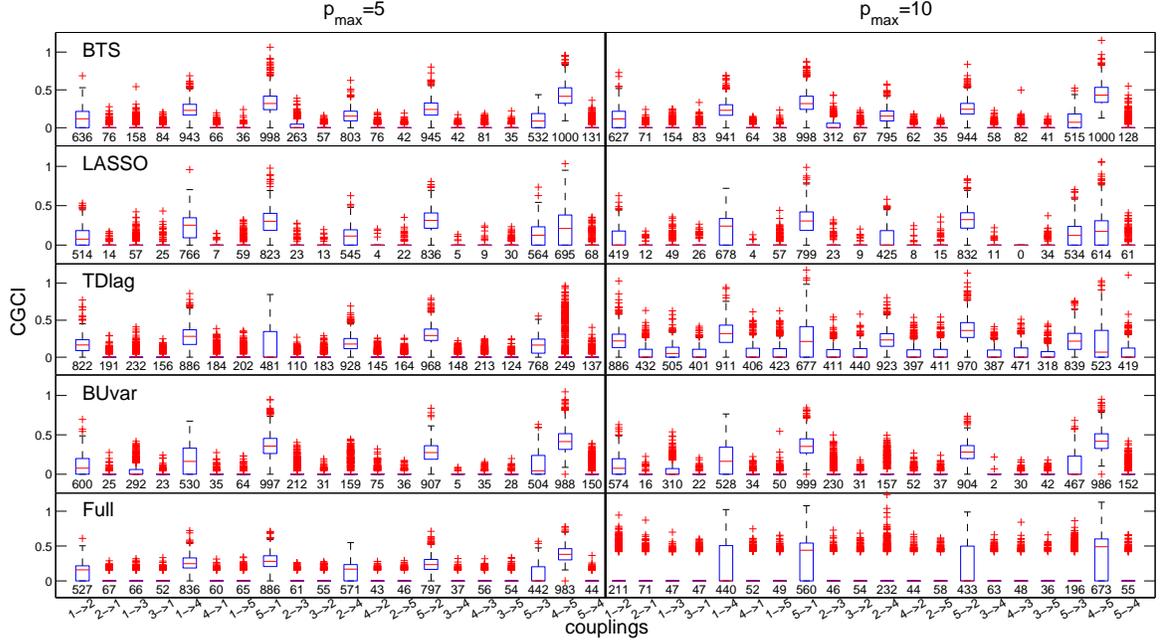}}
\caption{Boxplots of CGCI for all variable pairs of S1 and
$N=100$, from 1000 realizations,for
$p_{\mbox{\footnotesize{max}}}=5$ at the left panels and
$p_{\mbox{\footnotesize{max}}}=10$ at the right panels, and for
the methods BTS, LASSO, TDlag, BUvar and Full going from top to
bottom panels. At each panel the number of statistically
significant CGCI is displayed below each boxplot. The true
causality effects are: $X_1 \rightarrow X_2$, $X_1 \rightarrow
X_4$, $X_5 \rightarrow X_1$, $X_2 \rightarrow X_4$, $X_5
\rightarrow X_2$, $X_5 \rightarrow X_3$ and $X_4 \rightarrow
X_5$.} \label{fig:boxplots}
\end{figure*}

The rate of detection of the true Granger causality effects varies
with the method and the order, as shown in
Figure~\ref{fig:boxplots} for $p_{\mbox{\footnotesize{max}}}=5$
and $p_{\mbox{\footnotesize{max}}}=10$.
%----------------
%FIGURE(3)
%----------------
% ------------------------------------------------------------------------------------------------------------
The boxplots display the distribution of CGCI over the 1000
realizations for each ordered pair of variables and the number
below each boxplot is the number of times CGCI was found
statistically significant at $\alpha=0.05$ using the FDR
correction.

For $p_{\mbox{\footnotesize{max}}}=5$ being only one larger than
the correct order, LASSO, BTS and Full identify the seven true
causality effects at large percentages of the 1000 realizations,
with BTS scoring highest. The largest difference is observed for
$X_2 \rightarrow X_4$ and the detection percentage of BTS is
80.3\% to be compared with 54.5\% for LASSO and 57.1\% for Full.
However, BTS detects at a significant rate also false causality
effects (e.g. detection percentage 26.3\% for $X_2 \rightarrow
X_3$ as shown in Figure~\ref{fig:boxplots}), whereas LASSO and
Full do not exceed the significance level of 5\%. TDlag fails to
detect the coupling $X_4 \rightarrow X_5$ (detection percentage
24.9\%) and BUvar the coupling $X_1 \rightarrow X_3$ (detection
percentage 29.2\%), whereas they both give false causality effects
at a significantly high rate. These results suggest that the
bottom-up and top-down strategies are influenced by the order in
which the variables enter or leave the DR model, respectively.

When $p_{\mbox{\footnotesize{max}}}$ increases to 10, BTS and
BUvar are the most stable giving the same results, whereas LASSO
tends to detect Granger causality effects at a somewhat smaller
rate, Full does the same at a larger extent so that three true
couplings are detected at a rate smaller than 25\%, and TDlag
increases the detection rate of all couplings.

For each realization we calculate the measures sensitivity,
specificity, MCC, F-measure and Hamming distance on the basis of
the significance of CGCI for all pairs of variables of S1. We also
statistically compare the measures for mBTS and each of the other
methods. In Table~\ref{tab:System1}, the results on these
performance indices and the restriction methods, as well as Full,
are shown for $p_{\mbox{\footnotesize{max}}}=5$ and
$p_{\mbox{\footnotesize{max}}}=10$ and $N=100$.
%----------------
%TABLE(2)
%----------------
\begin{table*}
\renewcommand{\arraystretch}{1.3}
\caption{Sensitivity (SENS), specificity (SPEC), MCC, F-measure
(FM) and Hamming distance (HD) (average value and standard
deviation in parentheses) of the causality methods for S1,
$p_{\mbox{\footnotesize{max}}}=5,10$ and $N$=100. The highest
score at each performance index and
$p_{\mbox{\footnotesize{max}}}$ is highlighted.The statistical
significance for the mean difference of a measure on BTS and
another method is denoted with a superscript symbol
'\textsuperscript{+}' when the difference is positive and
'\textsuperscript{-}' when it is negative.} \centering
\bigskip
\label{tab:System1}
\renewcommand{\arraystretch}{1}
\begin{tabular}{c|c||c|c|c|c|c|c|c}
\hline
            & $p$  &    BTS  &   LASSO  & BUlag      & BUvar        &   TDlag       & TDvar     & Full  \\
\hline SENS & 5  & 0.823   &  0.673\textsuperscript{+} & 0.421\textsuperscript{+}   & 0.656\textsuperscript{+}     & 0.708\textsuperscript{+}      & {\bf 0.937}\textsuperscript{-}  & 0.556\textsuperscript{+}    \\
        (SD)       &    & (0.126) &  (0.171) & (0.167) & (0.139)   & (0.145)    &(0.087) & (0.184)   \\
                   & 10 & 0.819   & 0.612\textsuperscript{+}    & 0.414\textsuperscript{+}   & 0.650\textsuperscript{+}     & 0.794\textsuperscript{+}     & {\bf 0.953}\textsuperscript{-}   &  0.156\textsuperscript{+}       \\
               &    & (0.125) & (0.178)  & (0.167) & (0.138)   & (0.36)   & (0.080) &  (0.143)        \\
\hline SPEC & 5  & 0.935   & 0.977\textsuperscript{-}    & 0.950\textsuperscript{-}    & 0.938    & 0.880\textsuperscript{+}     & 0.762\textsuperscript{+}   & {\bf 0.984}\textsuperscript{-}      \\
        (SD)       &    &(0.073)  & (0.041)  & (0.061) & (0.064)   & (0.103)   & (0.143) & (0.037)           \\
                   & 10 & 0.934   & 0.979\textsuperscript{-}    & 0.954\textsuperscript{-}   & 0.935     & 0.651\textsuperscript{+}     & 0.401\textsuperscript{+}  & {\bf 0.990}\textsuperscript{-}         \\
               &    & (0.071) & (0.040)  & (0.058) & (0.066)   & (0.167)   &(0.168) & (0.033)         \\
 \hline MCC        & 5  & {\bf 0.775}   & 0.717\textsuperscript{+}    & 0.461\textsuperscript{+}   & 0.643\textsuperscript{+}     & 0.611\textsuperscript{+}      & 0.681\textsuperscript{+}  & 0.637\textsuperscript{+}    \\
        (SD)       &    & (0.155) & (0.154)  & (0.201) & (0.159)   & (0.177)    & (0.158) & (0.154) \\
                   & 10 & {\bf 0.746}   & 0.672\textsuperscript{+}    & 0.463\textsuperscript{+}   & 0.634\textsuperscript{+}     & 0.439\textsuperscript{+}      & 0.379\textsuperscript{+}  & 0.248\textsuperscript{+}    \\
               &    & (0.166) & (0.161)  & (0.191) & (0.163)   & (0.184)    & (0.162)& (0.197)   \\
 \hline FM  & 5  & {\bf 0.846}   & 0.773\textsuperscript{+}    & 0.542\textsuperscript{+}   & 0.736\textsuperscript{+}     & 0.732\textsuperscript{+}      & 0.796\textsuperscript{+}  & 0.683\textsuperscript{+}   \\
        (SD)       &    & (0.104) & (0.135)  & (0.173) & (0.116)   & (0.118)    & (0.096)& (0.160)  \\
                   & 10 & {\bf 0.843}   & 0.727\textsuperscript{+}    & 0.537\textsuperscript{+}   & 0.730\textsuperscript{+}     & 0.655\textsuperscript{+}      & 0.627\textsuperscript{+}  & 0.240\textsuperscript{+}    \\
               &    & (0.105) & (0.151)  & (0.0.171) & (0.119)   & (0.100)    & (0.073)& (0.199)  \\
 \hline HD         & 5  & {\bf 2.084}   & 2.587\textsuperscript{+}    & 4.707\textsuperscript{+}    & 3.217\textsuperscript{+}      & 3.608\textsuperscript{+}       & 3.528\textsuperscript{+}    & 3.321\textsuperscript{+}    \\
        (SD)       &    & (1.428) & (1.342)  & (1.509) & (1.373)   & (1.650)    & (1.920) & (1.312)  \\
                   & 10 & {\bf 2.123}   & 2.995\textsuperscript{+}    & 4.702\textsuperscript{+}   & 3.295\textsuperscript{+}      & 5.978\textsuperscript{+}     & 8.120\textsuperscript{+}   &  6.032\textsuperscript{+}   \\
               &    & (1.430) & (1.385)  & (1.437) & (1.411)    & (2.148)   & (2.205) &  0.972   \\
   \hline
   \end{tabular}
   \end{table*}
% =====================================================================================
BTS has high sensitivity and specificity, though it does not score
highest in both. The highest sensitivity score is obtained by
TDvar at 0.937 for $p_{\mbox{\footnotesize{max}}}=5$ and 0.953 for
$p_{\mbox{\footnotesize{max}}}=10$, both being statistically
significantly higher than for BTS, but TDvar scores lowest of all
methods in specificity. Similarly, Full scores highest in
specificity at 0.984 for $p_{\mbox{\footnotesize{max}}}=5$ and
0.990 for $p_{\mbox{\footnotesize{max}}}=10$, both significantly
higher than for BTS, but has the lowest sensitivity, which for
$p_{\mbox{\footnotesize{max}}}=10$ is remarkably low at 0.156. The
latter indicates the inappropriateness of the standard Granger
causality estimated by the full VAR model when the order is large.
The high sensitivity and specificity of BTS makes it score highest
in all three performance indices combining sensitivity and
specificity, and the difference is statistically significant for
all methods and for all three indices. Second best in MCC, FM and
Hamming distance is LASSO giving somewhat higher specificity than
BTS but much lower sensitivity (for FM TDvar scores slightly
higher than LASSO when $p_{\mbox{\footnotesize{max}}}=5$ and BUvar
when $p_{\mbox{\footnotesize{max}}}=10$). All the methods
restricting the VAR model are robust to the increase of
$p_{\mbox{\footnotesize{max}}}$ and give approximately the same
performance index values, while the performance of Full drops
drastically, e.g. the number of false positives and negatives
counted by the Hamming distance are doubled (from about 3 to 6)
going from $p_{\mbox{\footnotesize{max}}}=5$ to
$p_{\mbox{\footnotesize{max}}}=10$. The bottom-up strategies BUlag
and BUvar have as high specificity as BTS but much lower
sensitivity, especially BUlag. Thus for this system, the time-lag
supervised bottom-up search in BTS manages to spot the correct
lagged variables better than an arbitrary bottom-up search of
BUlag and BUvar. The results for all methods vary across
realizations but at about the same amount, which is larger for the
sensitivity than for the specificity, as indicated by the
standard deviation (SD) values. Full and LASSO
tend to have the largest SD of sensitivity and smallest SD of
specificity, which amounts the same SD of MCC as for BTS but
larger SD of the FM than for BTS. We note that the SD is
moderately large with respect to the average values but allows for
the mean differences to be statistically significant.

The results in Table~\ref{tab:System1} are obtained using the
significance test for CGCI corrected with FDR. If no correction
for multiple testing is applied the sensitivity increases at the
cost of lower specificity and the overall performance measured by
MCC, F-measure and Hamming distance is somewhat lower. If no
significance testing is performed, and the existence of coupling
is determined simply by a zero or positive CGCI (this is not
applied to Full), the sensitivity is further increased but the
specificity is disproportionately decreased, giving smaller
overall performance indices. For example, for BTS and
$p_{\mbox{\footnotesize{max}}}=5$, the sensitivity increases from
0.823 to 0.867 but the specificity drops from 0.935 to 0.821, so
that MCC drops from 0.775 to 0.674. The same feature is observed
for all but LASSO methods restricting VAR, so that the superiority
of BTS is maintained also when significance testing is not
performed, and it remains superior to Full, which is always
applied with the significance test. It is noted that LASSO cannot
be included in this comparison as it contains a significance test
for the model coefficients (the covariance test, see
Sec.~\ref{sec:OtherMethods}), and thus it gives about the same
results with and without the significance test for CGCI.

% -------------------------------
\subsection{System 2}
% -------------------------------

In Table~\ref{tab:System2}, the average sensitivity, specificity
and MCC from 1000 realizations
 are shown for system S2 using as maximum order the
true VAR order ($p_{\mbox{\footnotesize{max}}}=5$) and different
time series lengths $N=50,100,1000$.
%----------------
% TABLE(3)
%----------------
\begin{table*}
\caption{Average sensitivity (SENS), specificity (SPEC) and MCC of
the causality methods for system S2,
$p_{\mbox{\footnotesize{max}}}=5$ and $N=50,100,1000$. The highest
score at each performance index and $N$ is highlighted. The
statistical significance for the mean difference of a measure on
BTS and another method is denoted with a superscript symbol
'\textsuperscript{+}' when the difference is positive and
'\textsuperscript{-}' when it is negative.} \centering
\bigskip
\label{tab:System2}
\renewcommand{\arraystretch}{1}
\begin{tabular}{c|c||c|c|c|c|c|c|c}
\hline
              & $N$     &   BTS         &   LASSO  & BUlag      & BUvar        &   TDlag       & TDvar     & Full  \\
\hline SENS   & 50      & 0.916        & 0.774\textsuperscript{+}    & 0.857\textsuperscript{+}      & 0.848\textsuperscript{+}        & 0.964\textsuperscript{-}         & {\bf 0.985}\textsuperscript{-}     & 0.727\textsuperscript{+}   \\
              & 100     &  0.996         & 0.919\textsuperscript{+}     & 0.939\textsuperscript{+}       & 0.935\textsuperscript{+}         &  0.996         & {\bf 1.0}\textsuperscript{-}          & 0.993    \\
              & 1000    & {\bf 1.0}             & 0.999    & {\bf 1.0}    & {\bf 1.0}            & {\bf 1.0}             & {\bf 1.0}         & {\bf 1.0}    \\
 \hline SPEC  & 50      &  0.947       & {\bf 0.976 }\textsuperscript{-}     & 0.958\textsuperscript{-}       & 0.962\textsuperscript{-}         & 0.848\textsuperscript{+}          & 0.747\textsuperscript{+}      & {\bf 0.976}\textsuperscript{-}    \\
              & 100     & 0.967         & {\bf 0.978}\textsuperscript{-}     & 0.977\textsuperscript{-}       & 0.974\textsuperscript{-}         & 0.921\textsuperscript{+}          & 0.826\textsuperscript{+}      &  0.973    \\
              & 1000    & 0.987         & 0.981\textsuperscript{+}     & {\bf 0.993}\textsuperscript{-} & 0.992\textsuperscript{-}         & 0.982\textsuperscript{+}          & 0.865\textsuperscript{+}      & 0.976\textsuperscript{+}     \\
 \hline MCC   & 50      & {\bf 0.868}   & 0.799\textsuperscript{+}     & 0.838\textsuperscript{+}       & 0.836\textsuperscript{+}         & 0.796 \textsuperscript{+}         & 0.713\textsuperscript{+}      & 0.762\textsuperscript{+}     \\
              & 100     & 0.955         & 0.910\textsuperscript{+}     & 0.924\textsuperscript{+}       & 0.918\textsuperscript{+}         & 0.901\textsuperscript{+}          & 0.803\textsuperscript{+}      & {\bf 0.961}    \\
              & 1000    &  0.983          & 0.975\textsuperscript{+}     & {\bf 0.991}\textsuperscript{-}  & 0.989\textsuperscript{-}         & 0.977\textsuperscript{+}          & 0.843\textsuperscript{+}      & 0.969\textsuperscript{+}     \\
   \hline
   \end{tabular}
   \end{table*}
% ----------------------------------------------------------------------
For $N=50$ and $100$ the highest sensitivity score is obtained by
TDvar, at 0.985 and 1.0 respectively, but this method scores
lowest in specificity. For $N=1000$, all methods detect the true
couplings and score 1.0 in sensitivity (almost 1.0 for LASSO). For
$N=50$, the highest specificity score is obtained by LASSO and
Full but both have the lowest score in sensitivity. When $N$
increases, the specificity, as well as the sensitivity and MCC,
approaches 1.0 for all but TDvar methods. BTS method may not rank
first in sensitivity and specificity but consistently presents
high values for all time series lengths, so that it scores highest
in MCC for $N=50$ and follows closely with the highest MCC for
larger $N$. It is also noted that BTS scores higher in MCC than
LASSO for all time series lengths. All the mean differences in the
indices between BTS and any of the other measures are found
statistically significant (see the plus and minus superscripts in
Table~\ref{tab:System2}).

% -------------------------------
\subsection{System 3}
% -------------------------------

System 3 is VAR on 20 variables and order three and has 10\%
non-zero coefficients of a total of 1200 coefficients, giving
respectively 10\% true couplings of a total of 380 possible
ordered couplings. Considering the DR model for each of the 20
variables, the total number of coefficients is 60 and thus Full
cannot provide stable solution when $N$ is as small as 100, being
unable to identify Granger causality effects and giving therefore
a very small sensitivity value, 0.002, as shown in
Table~\ref{tab:System3}.
%----------------
%TABLE(4)
%----------------
\begin{table*}
\caption{Average sensitivity (SENS), specificity (SPEC) and MCC of
the causality methods for S3, and combinations of
$p_{\mbox{\footnotesize{max}}}=4,5$ and $N=100,200,500$.The
statistical significance for the mean difference of a measure on
BTS and another method is denoted with a superscript symbol
'\textsuperscript{+}' when the difference is positive and
'\textsuperscript{-}' when it is negative.} \centering
\bigskip
\label{tab:System3}
\renewcommand{\arraystretch}{1}
\begin{tabular}{c|c|c||c|c|c|c|c|c|c}
\hline
              &     $p$   & $N$       &   BTS         &  LASSO  & BUlag    & BUvar       & TDlag   &   TDvar  &  Full   \\
\hline SENS   &      4    & 100       &  0.238        & 0.091 \textsuperscript{+}    & 0.233    & 0.228\textsuperscript{+}        &  0.624\textsuperscript{-}    &  {\bf 0.690}\textsuperscript{-}   & 0.002\textsuperscript{+}    \\
              &      4    & 200       & 0.506         & 0.258\textsuperscript{+}    & 0.576\textsuperscript{-}     & 0.583\textsuperscript{-}  & 0.718\textsuperscript{-}    & {\bf 0.793}\textsuperscript{-}     & 0.064\textsuperscript{+}  \\
              &     5     & 500       & 0.959         & 0.791\textsuperscript{+}    & 0.949\textsuperscript{+}     & 0.964\textsuperscript{-}  & 0.978\textsuperscript{-}    & {\bf 0.993}\textsuperscript{-}     & 0.775\textsuperscript{+} \\
\hline SPEC   &     4     & 100       &  0.989        & 0.998\textsuperscript{-}    & 0.990    & 0.991       & 0.637\textsuperscript{+}    & 0.524\textsuperscript{+}     & {\bf 0.999 }\textsuperscript{-}    \\
              &    4      & 200       & 0.996         & {\bf 0.999}\textsuperscript{-}    & 0.992\textsuperscript{+}     & 0.992\textsuperscript{+}  & 0.945\textsuperscript{+}    & 0.834\textsuperscript{+}     & {\bf 0.999} \\
              &    5      & 500       & 0.992         & {\bf 0.997}\textsuperscript{-}    & 0.994\textsuperscript{+}     & 0.993\textsuperscript{+}  & 0.966\textsuperscript{+}    & 0.882\textsuperscript{+}     & 0.995\textsuperscript{-} \\
\hline  MCC   & 4         & 100       & 0.373         & 0.248\textsuperscript{+}    & 0.373    & {\bf 0.376}\textsuperscript{-}  & 0.164\textsuperscript{+}    & 0.131\textsuperscript{+}     & 0.004\textsuperscript{+}    \\
              &   4       & 200       & 0.664         & 0.470\textsuperscript{+}    & 0.693\textsuperscript{-}     & {\bf 0.695}\textsuperscript{-}  & 0.616\textsuperscript{+}    & 0.457\textsuperscript{+}     & 0.198\textsuperscript{+} \\
              &     5     & 500       & 0.941         & 0.788\textsuperscript{+}    & 0.942   & {\bf 0.948}\textsuperscript{-}  & 0.942    & 0.659\textsuperscript{+}     & 0.845\textsuperscript{+} \\
\hline
   \hline
   \end{tabular}
   \end{table*}
% ===================================================================================
Neither for $N=200$, Full identifies the true Granger causality
effects (sensitivity at 0.064), and it achieves this when $N=500$,
scoring still lowest of all methods. The highest score in
sensitivity is obtained by TDvar for all $N$, being significantly
higher than for BTS, but again it scores lowest in specificity for
all $N$. Highest in specificity is again Full competing with LASSO
for the first place, but all but the top-down strategies score
almost equally high, still being both significantly higher than
for BTS. LASSO scores much lower in sensitivity so that overall it
performs purely in this high-dimensional system. The best
performance, as quantified by MCC, is again exhibited by the
bottom-up strategies, with BTS scoring slightly below the largest
MCC obtained by BUvar, but this difference is found statistically
significant. The other methods (LASSO, TDlag and TDvar, Full) do
much worse and do not approach as $N$ increases the level of
bottom-up strategies. For example for $N=500$, the bottom-up
strategies give MCC about 0.94, whereas the other methods score up
to 0.85.

% -------------------------------
\subsection{System 4}
% -------------------------------

\begin{figure*}[htb]
\hbox{\centerline{\includegraphics[width=7cm]{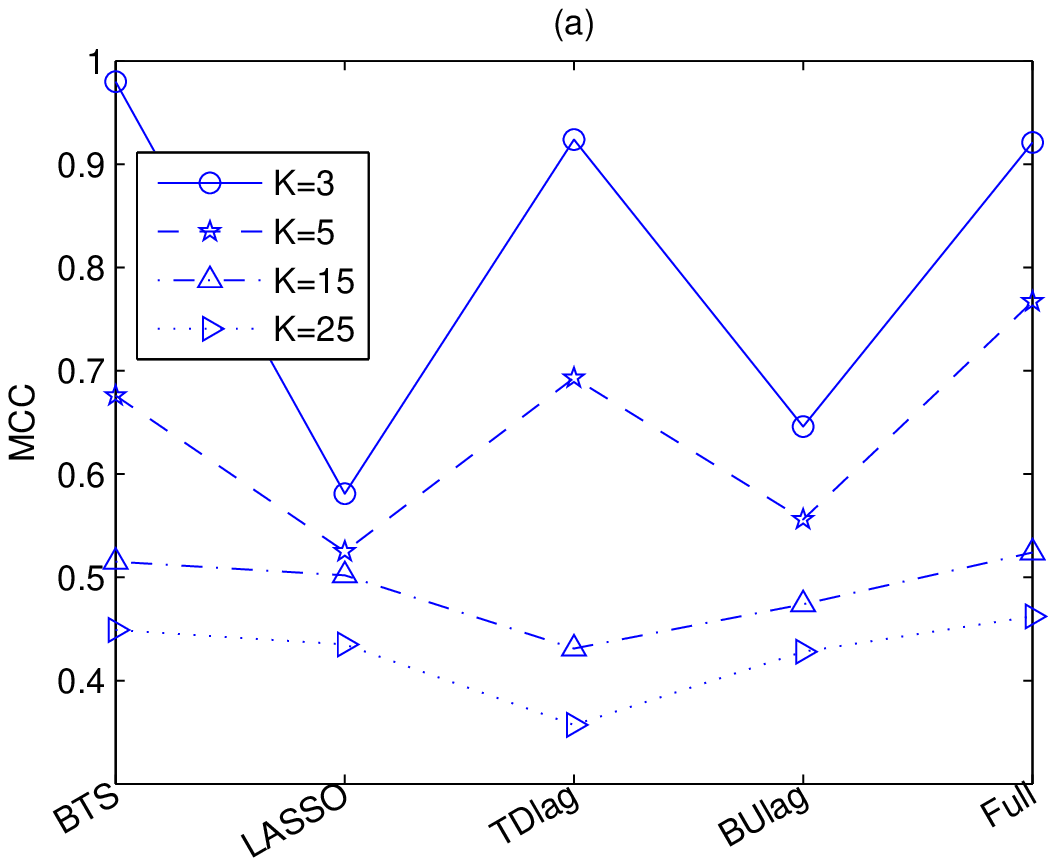}
\includegraphics[width=7cm]{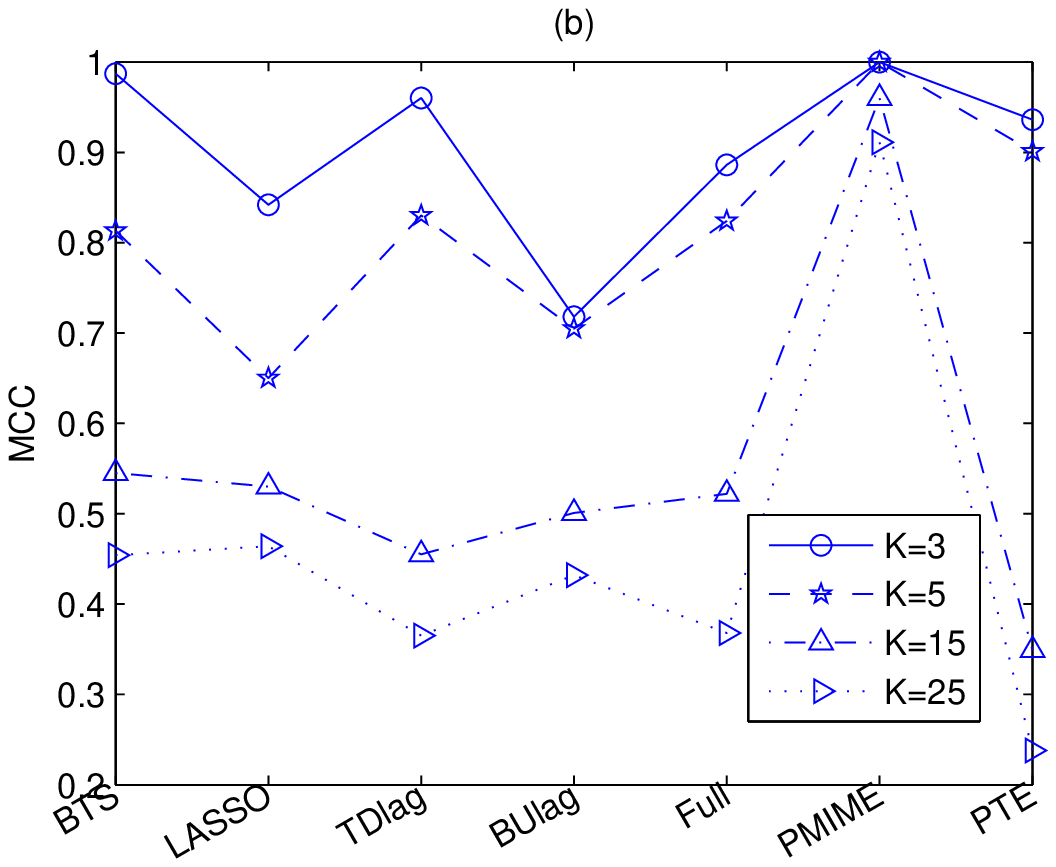}}}
\caption{MCC vs method as given in the x-axis for S5 (coupled
H\'{e}non maps), $p_{\mbox{\footnotesize{max}}}=5$ and number of
variables $K$ as given in the legend. (a) $N=512$, (b) $N=1024$,
including the nonlinear methods PMIME and PTE.}
\label{fig:MCCHenon}
\end{figure*}

Many real world problems may involve nonlinear causality
relationships and it is therefore of interest to assess at what
extent the linear Granger causality methods can estimate correctly
nonlinear causal effects. Certainly, for a comprehensive
assessment by means of a simulation study one should include
systems having different forms of nonlinear relationships, but
here we restrict ourselves to system S4, a known testbed for
nonlinear Granger causality methods. We consider in the comparison
two nonlinear causality methods based on information theory, the
partial transfer entropy (PTE) computed on embedding vectors from
each variable corresponding to CGCI on the full VAR representation
\cite{Vakorin09,Papana12}, and the partial mutual information on
mixed embedding (PMIME) computed on restricted mixed embedding
vectors from all variables corresponding to CGCI on the restricted
VAR representation \cite{Vlachos10,Kugiumtzis13a}. PTE is the
extension of the widely used bivariate transfer entropy (TE)
\cite{Schreiber00}, taking into account the presence of the other
observed variables. PMIME first searches for the most relevant
lagged variables to the response $X_j$ using the conditional
mutual information. Then similarly to the definition of PTE, PMIME
quantifies the information of the lagged variables of the driving
variable $X_i$ to the response variable $X_j$. For the entropy
estimation in both PTE and PMIME the estimate of $k$-nearest
neighbors is used \cite{Kraskov04}. Figure~\ref{fig:MCCHenon}
shows the value of MCC for different number of variables $K$ and
for time series lengths $N=512, 1024$, for the latter reporting
also results for PMIME and PTE (obtained from 100 realizations as
opposed to 1000 realizations used for the linear methods).
%----------------
% Figure(4)
%----------------
% ---------------------------------------------------------------------
We observe that among the linear methods, for small $K$ BTS is the
best followed by TDlag and LASSO is the worst, whereas for larger
$K$ BTS is again the best followed closely by LASSO and the worst
is TDlag. BUlag performs similarly to LASSO and Full similarly to
BTS scoring highest for $N=512$ and $K=5$. Full still scores high
for large $K$ and all linear methods have reduced specificity as
$K$ increases scoring low in MCC. This is to be contrasted to
PMIME for $N=1024$ obtaining MCC$>$0.9 even for $K=25$, whereas
PTE fails when $K$ gets large ($K=15,25$), because it does not
apply any restriction to the lagged variables.

Dealing with systems of many variables, high order and long time
series, the computation cost may be an issue, and it is known that
a main shortcoming of LASSO is the computation time. It is not
easy to give exact figures for the computational complexity, as it
depends on the sparsity of the matrix of the lagged variables of
the true system. The computational complexity can be measured in
terms of the tested models, and for BTS it is a multiple of the
number $K$ of variables in the model, comparable (slightly larger)
to that for the bottom-up and top-down strategies. To the
contrary, the computational time of LASSO is much larger. As
mentioned above, LASSO is computed in two parts: the different
models derived for increasing tuning parameter $s$ ('lars'
function) and the selection of the final model using the
covariance test ('covTest' function). We counted the computation
times for the two parts of LASSO in the simulations for different
systems, time series lengths and orders and we concluded that the
second part is the most time consuming. For example, for one
realization of the S1 system ($N=100$,
$p_{\mbox{\footnotesize{max}}}=10$, $K=5$) and for all $K$
variables, LASSO was completed in 2.81 s (first part: 0.77 s,
second part: 2.04 s) while BTS in 0.10 s \footnote{The
computations were done in a PC of 3.16GHz CPU, Core2Duo, 4Gb
RAM.}. Also, for one realization of the S3 system ($N=500$,
$p_{\mbox{\footnotesize{max}}}=5$, $K=20$), LASSO was completed in
83.46 s (first part: 1.22 s, second part: 82.236 s) while BTS in
2.59 s. Overall, in the simulation study LASSO was found to be
about 20 times slower than BTS, which in turn is more than 5 times
slower than the top-down and bottom-up strategies, the latter
giving about the same computation times.

%%%%%%%%%%%%%%%%%%%%%%%%%%%%%%%%%%%%%%%%%%%%%%%%%%%%%%%%%%%%%%%%%%%%%%%
\section{Application to real data}
\label{sec:RealData}
%%%%%%%%%%%%%%%%%%%%%%%%%%%%%%%%%%%%%%%%%%%%%%%%%%%%%%%%%%%%%%%%%%%%%%%

In this application we assess whether the VAR restriction allows
for a better quantification of the Granger causality effects
between brain areas, known also as effective connectivity
\cite{Friston94}. The dataset is a scalp multi-channel
electroencephalographic (EEG) recording of a patient with epilepsy
containing 8 episodes of epileptiform discharge (ED), i.e. a small
electrographic seizure of very small duration. The recording was
done at the Laboratory of Clinical Neurophysiology, Medical
School, Aristotle University of Thessaloniki. The EEG data were
initially sampled at 1450 Hz and downsampled to 200 Hz,  band-pass
filtered at [0.3,70] Hz (FIR filter of order 500), initially
referenced to the right mastoid and re-referenced to infinity.
Channels with artifacts were removed resulting in 44 artifact-free
channels. For each of the 8 episodes, the ED is terminated by the
administration of transcranial magnetic stimulation (TMS) (a block
of 5 TMS at 5 Hz frequency). For details on the experimental setup
see \cite{Kimiskidis13,Kugiumtzis15a}. For all episodes, we
considered a data window of about 23 sec including a part before
the ED, the ED and a part after the ED. Each data window was split
to overlapping sliding windows of duration 2 sec and sliding step
1 sec. We consider three states for each episode, the preED state
regarding the first 9 windows before the ED start, the ED state
covering the ED (3-5 windows, depending on the episode), and the
postED state regarding the time after the end. The interest is in
discriminating the three states on the basis of the connectivity
(causality) structure of the EEG channels.

The brain (effective) connectivity structure can be better seen in
a network form, where the nodes of the network are the EEG
channels and the weighted connections of the network are given by
the corresponding CGCI values. Certainly, we cannot anticipate the
detection of specific causality effects between any channels, but
we expect that the overall connectivity is different during ED
than before and after ED. The brain connectivity obtained by CGCI
can be quantified using network measures, and here we summarize
the connectivity of each channel by the out-strength of the node
at each sliding window, $s_{i}= \frac{1}{K-1}\sum_{j=1}^K
\mbox{CGCI}_{i\rightarrow j}$, the total driving effect of channel
$i$ to all other $K$ channels, and thereafter we compute the
average strength over all nodes, $S=\frac{1}{K}\sum_{i=1}^K
s_{i}$.

\begin{figure*}[htb]
\centerline{\hbox{\includegraphics[height=60mm]{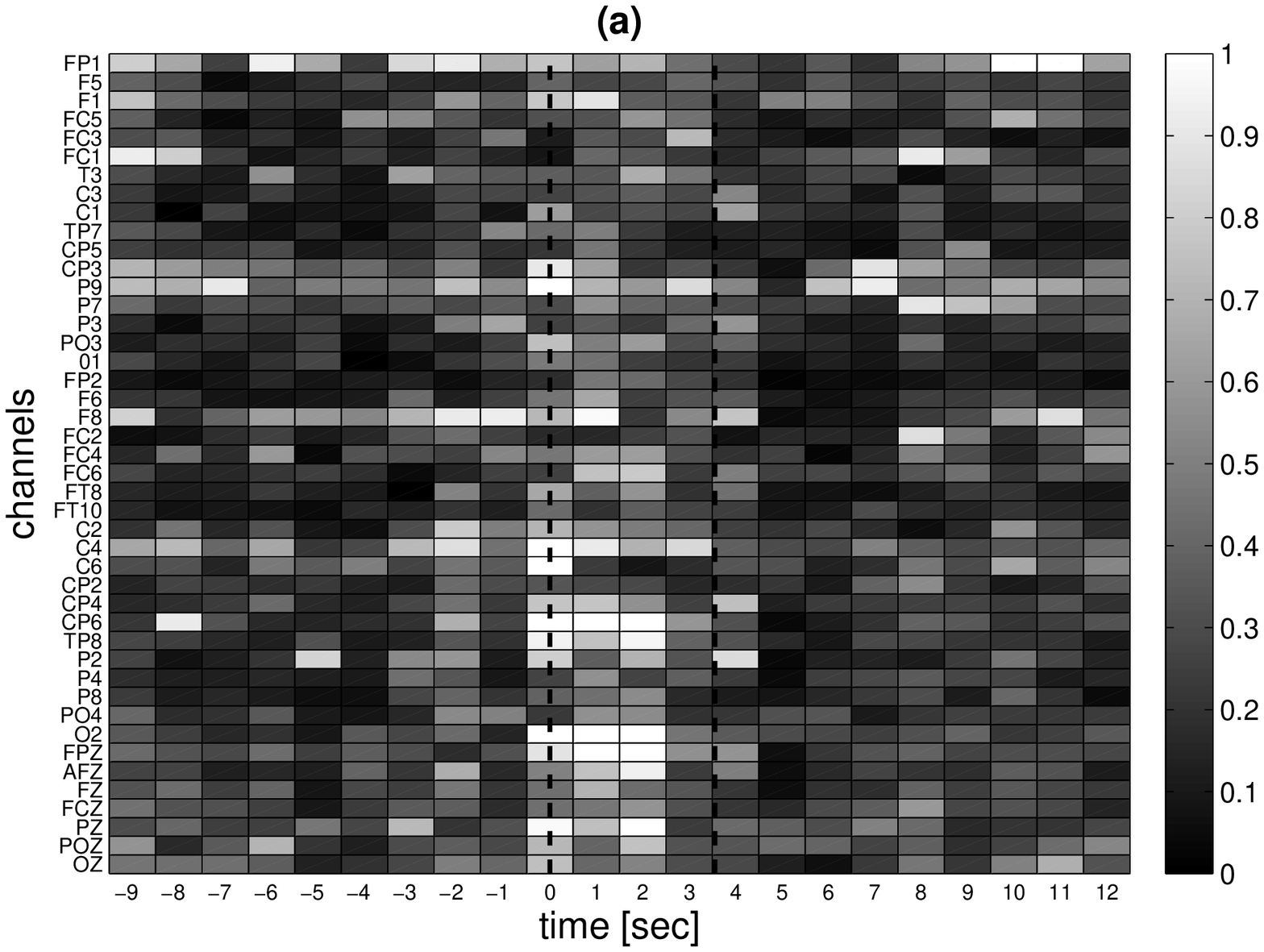}
\includegraphics[height=60mm]{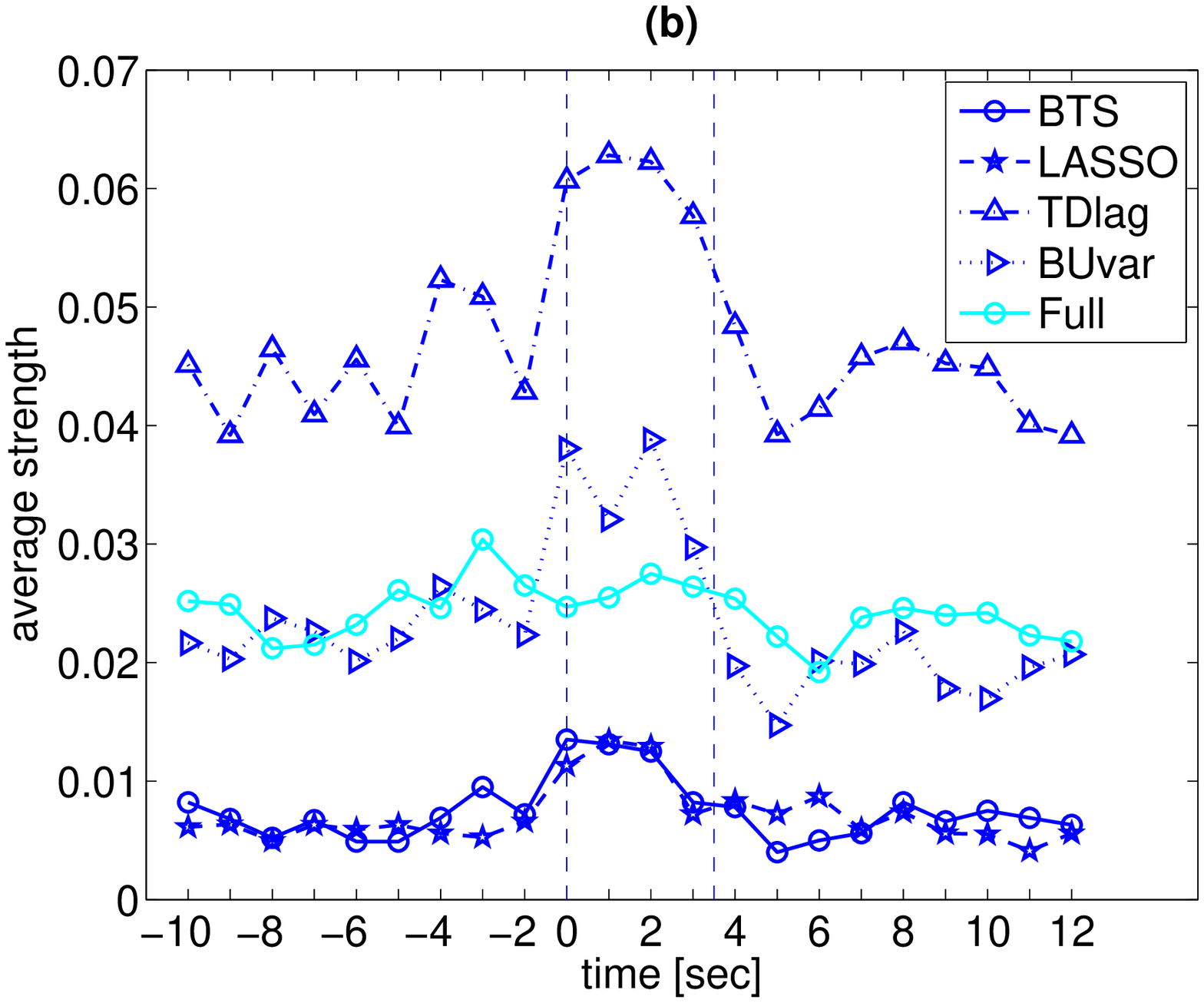}}}
\caption{(a) Color map of node (EEG channel) strength with BTS
method in successive windows across one ED episode for
$p_{\mbox{\footnotesize{max}}}=3$ and $N=400$. (b) The average
strength for the VAR restriction methods and Full, as shown in the
legend, for successive sliding windows and the same episode as in
(a). The dashed lines indicate the start and end of ED and the
time is given with reference to the ED start.} \label{fig:realBTS}
\end{figure*}

CGCI derived by BTS exhibits a clear increase in connectivity
during ED, whereas there are few causal effects at the preED and
postED state. An example of the $s_i$ across all time windows of
one episode is shown in Figure~\ref{fig:realBTS}a.

%----------------
%Figure(5)
%----------------
% ----------------------------------------------------------------------

The increased connectivity at the ED state is observed mostly at
the frontal and central channels (channel names starting with
``F'' or ``C'' at the lower part of Figure~\ref{fig:realBTS}a),
whereas the occipital channels (names starting with ``O'') exhibit
almost no connectivity during ED and some low-level connectivity
during preED and postED.

The $S$ as a function of the time window for the same episode is
shown in Figure~\ref{fig:realBTS}b for the VAR restriction methods
BTS, LASSO, TDlag, BUvar, as well as Full. It is clearly shown
that $S$ increases during ED for the VAR restriction methods but
not for Full, for which $S$ fluctuates regardless of the state. We
also tested larger $p_{\mbox{\footnotesize{max}}}$ values, but the
difference between preED, ED and postED was depicted better for
$p_{\mbox{\footnotesize{max}}}=3$. The differences in the range of
strength across different methods, as shown in
Figure~\ref{fig:realBTS}b, is due to the different number of
variables (EEG channels) selected by the methods in the dynamic
regression model. For example, in the DR of $X_{40}$, 12 variables
are found by BTS, 7 by LASSO, 27 by TDlag and 16 by BUvar. As the
number of variables increases in the DR so does the number of
causal effects and subsequently the connections in the network.
Despite this difference, all restriction methods reveal well the
difference between the states.

\begin{figure*}[htb]
\hbox{\centerline{\includegraphics[width=7cm]{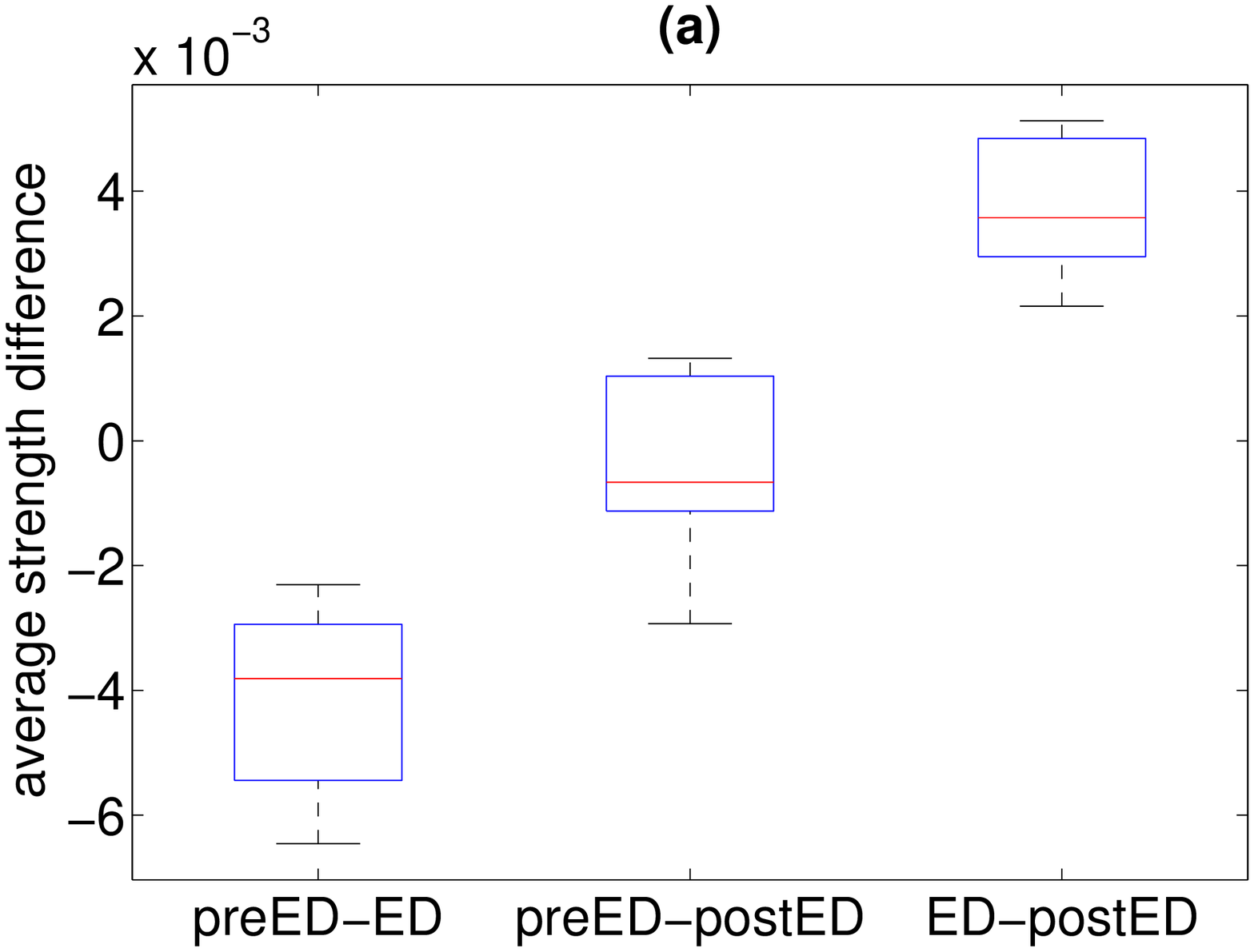}
\includegraphics[width=7cm]{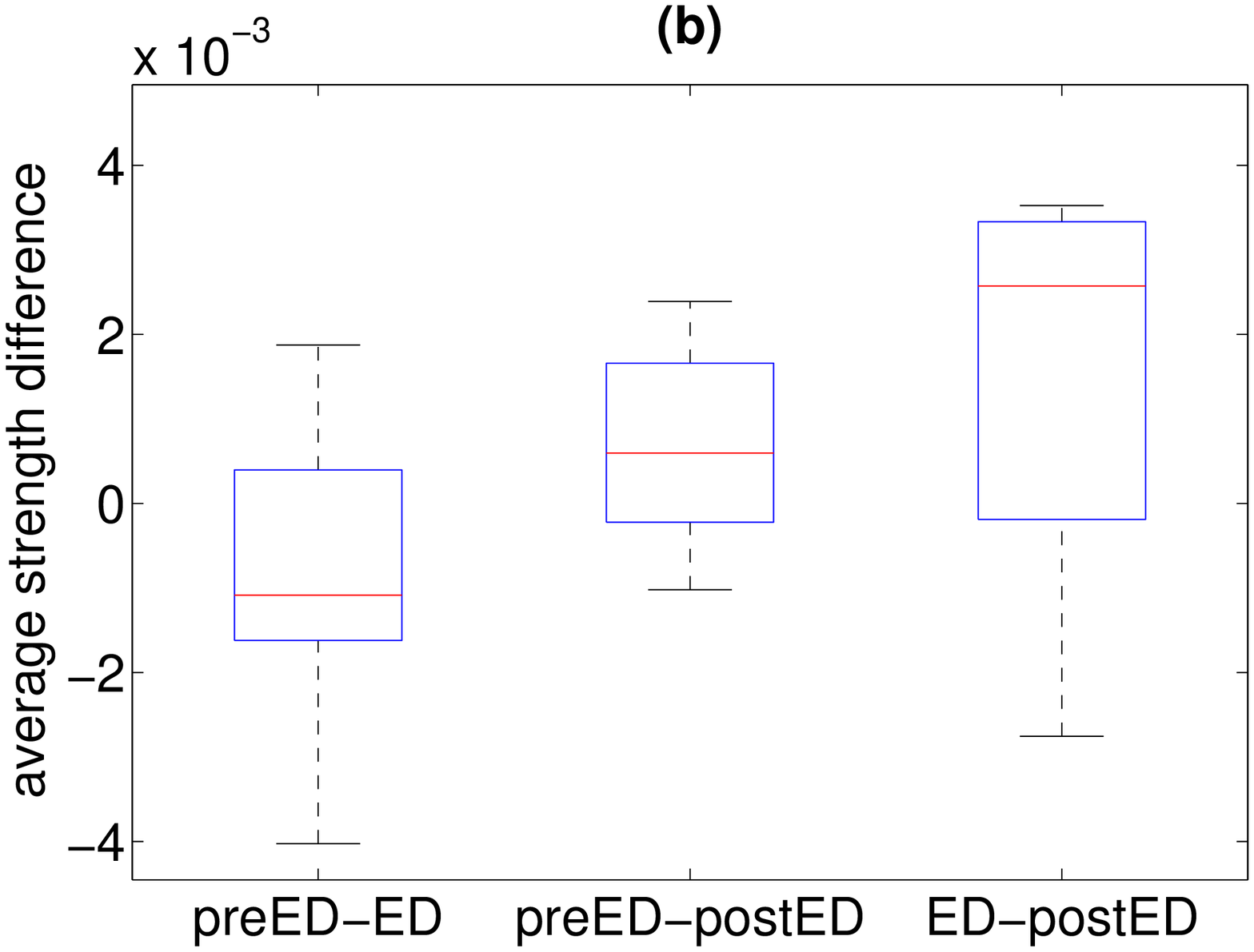}}}
\caption{Boxplots for the average strength differences of
preED-ED, preED-postED and ED-posteED from 8 episodes:(a) BTS and
$p_{\mbox{\footnotesize{max}}}=3$,(b) Full and
$p_{\mbox{\footnotesize{max}}}=3$.} \label{fig:realS}
\end{figure*}

To assess the statistical significance of the change of
connectivity from the preED state to the ED state and back to the
postED states, we applied paired samples student test for each of
the pair differences preED-ED, postED-ED and preED-postED state,
where the sample for each state contains 8 observations, and each
observation is the average of the network strength $S$ over all
windows for the corresponding state in an episode. In
Figure~\ref{fig:realS}, the boxplots for the three pair
differences in $S$ derived by BTS and Full are presented.
%----------------
% Figure(6)
%----------------
% ----------------------------------------------------------------------
For BTS, the mean differences preED-ED and ED-postED are found
statistically significant ($p=0.00012$ and $p=0.00003$,
respectively) and the same was obtained by the Wilcoxon sign rank
test, whereas the difference preED-postED is not found
statistically significant. The same significant differences could
be established also with the other VAR restricting methods but
TDlag. On the other hand, for Full statistically significant
difference could not be established for any states. This finding
indicates the necessity of using VAR restriction methods when the
linear Granger causality index has to be computed on a large
number of EEG channels.

%%%%%%%%%%%%%%%%%%%%%%%%%%%%%%%%%%%%%%%%%%%%%%%%%%%%%%%%%%%%%%%%%%%%%%%
\section{Discussion}
\label{sec:Discussion}
%%%%%%%%%%%%%%%%%%%%%%%%%%%%%%%%%%%%%%%%%%%%%%%%%%%%%%%%%%%%%%%%%%%%%%%

We introduce a new approach in forming the classical conditional
Granger causality index (CGCI) for the estimation of the Granger
causality in time series of many variables, i.e. instead of
deriving CGCI on the full vector autoregressive (VAR)
representation, we suggest a restriction of VAR using the method
of backward-in-time selection (BTS). We call this approach
BTS-CGCI. BTS is a bottom-up strategy and it builds up the dynamic
regression model progressively searching first for lagged
variables temporally closer to the response variable.
 This supervised search is designed exactly for
 time series problems, unlike other methods, such as LASSO, being
 developed for regression problems.
We propose a modified BTS, which results in specific lagged
variables instead of all the lagged variables up to a selected
order. For the inclusion of a lagged variable in the model, the
Bayesian information criterion (BIC) is utilized. The same
criterion is used also in the implementation of the bottom-up and
top-down strategies, to which BTS is compared. It is noted that
the criterion of final prediction error (FPE) was used in the
simulations, and gave qualitatively similar results.

The use of BTS allows for a convenient deduction of lack of
Granger causality from CGCI. If the dynamic regression model
obtained by BTS does not contain any components of the driving
variable then CGCI is set to zero and there is no Granger
causality effect. Otherwise, this model is considered as the
unrestricted model (containing the driving variable components)
and the restricted model is derived by omitting the components of
the driving variable. Then CGCI is computed on the basis of these
two models, as in the standard definition of CGCI. Furthermore,
significance test on CGCI can be performed as for the CGCI with
the full VAR (correcting also for multiple testing). However, for
applications where parametric significance testing cannot be
trusted due to deviation from normality, a positive CGCI derived
from BTS can be considered as significant. The simulation study
showed that this simpler approach gains better sensitivity in
detecting causal effects than when using the significance test. On
the other hand, it tends to give more false positives so that
overall the implementation with the significance testing should be
preferred when it is applicable.

BTS was compared to other VAR restriction methods, i.e. bottom-up
and top-down strategies and LASSO, in their ability to identify
the correct causal relationships. The simulation results in the
cases of three VAR processes and a nonlinear coupled dynamical
system showed that BTS had the overall best performance scoring
consistently high in the performance indices of sensitivity,
specificity, Matthews correlation coefficient (MCC), F-measure
(FM) and Hamming distance. The other methods had varying
performance, and in particular LASSO scored generally low.
Differences among the methods were particularly revealed at small
time series lengths $N$, whereas for larger $N$ all methods tended
to converge in estimating the correct causal effects. However,
this was not true for the full VAR representation (the standard
approach) whenever the VAR order was large. For small $N$, LASSO
and the full VAR tended to have the highest specificity but
relatively low sensitivity, whereas the top-down strategy tended
to have the highest sensitivity but also the lowest specificity.
On the other hand, BTS scored always high in both sensitivity and
specificity, resulting in the highest or close to the highest
score in MCC, F-measure and Hamming distance in all settings. We
attribute the good performance of CGCI with BTS to the design of
BTS that unlike the other methods it was developed for time series
(not regression in general) making use of the time order of the
lagged variables.

Our simulation study confirmed the computational shortcoming of
LASSO of being much slower than all other methods. The top-down
strategy requires the estimation of the full model in the initial
step, thus it bears the same shortcoming as the full VAR for many
variables and short time series. On the other hand, the bottom-up
strategy and BTS start from few (actually none) degrees of freedom
adding one lagged variable at a time, and therefore they are both
computationally and operationally effective for short time series
even when the number of time series is very large. LASSO can also
be used in this setting but not in conjunction with the covariance
test that gives the significant lagged variable terms. Thus BTS
can be applied even in settings where the number of variables
exceeds the time series length, as for example in gene microarrays
experiments.

We confirmed the significance of BTS along with the other VAR
restriction methods in an application of epileptiform discharge
observed in electroencephalograms. We could clearly detect the
change of brain connectivity during the epileptiform discharge
with the VAR restricted methods that could not be revealed by the
full VAR.

In some applications, and particularly in econometrics, the
Granger causality has to be estimated in non-stationary time
series. If the time series are also co-integrated, the model of
choice is the vector error correction model (VECM) that contains a
full VAR model for the first differenced variables \cite{Tsay02}.
The dimension reduction, and mBTS in particular, can be extended
to restrict the VAR part in VECM, but the implementation is not
straightforward and is left for future work. If the time series
are not co-integrated then the non-stationarity for each time
series separately is first removed, e.g. taking first differences,
and the dimension reduction method can then be applied to the
stationary time series in a straightforward manner.

In summary, the main aspects of the proposed approach are:
\begin{itemize}
    \item The dimension reduction methods provide overall
    better estimation of Granger causality than the standard full
    VAR approach.
    \item The recently developed dimension reduction method of
    backward-in-time selection (BTS) has been modified (mBTS), so
    as to give the exact lagged variables being the most
    explanatory to the response variable.
    \item The proposed mBTS method uses a time ordered supervised
    sequential selection of the lagged variables, tailored for
    time series, as opposed to other methods that do not take into
    account the time order, being developed for regression
    problems (bottom-up, top-down, LASSO).
    \item The simulation study showed that mBTS estimates
    overall better the Granger causality
    than other tested dimension reduction methods (bottom-up, top-down,
    LASSO) and the standard VAR approach.
    \item The proposed mBTS method is particularly useful for
    high dimensional time series of short length, as demonstrated
    in the application to epileptic EEG.
\end{itemize}

The Matlab codes for mBTS and CGCI are given in the second
author's homepage, currently being {\tt
http://users.auth.gr/dkugiu/}.

%%%%%%%%%%%%%%%%%%%%%%%%%%%%%%%%%%%%%%%%%%%%%%%%%%%%%%%%%%%%%%%%%%%%%%%
\section{Acknowledgments}
%%%%%%%%%%%%%%%%%%%%%%%%%%%%%%%%%%%%%%%%%%%%%%%%%%%%%%%%%%%%%%%%%%%%%%%
The work is supported by the Greek General Secretariat for
Research and Technology (Aristeia II, No 4822). The authors want
to thank Vasilios Kimiskidis, at Aristotle University of
Thessaloniki, for providing the EEG data, and Ioannis Rekanos, at
Aristotle University of Thessaloniki for valuable comments and
discussions.

% \bibliography{c:/MyFiles/Papers/LaTeX/allref}
\bibliographystyle{IEEEtran}


\end{document}